\documentclass[prd,aps,nofootinbib]{revtex4}

\usepackage{amssymb,amsmath,epsfig,bm}
\usepackage[bookmarks]{hyperref}
\usepackage[usenames,dvipsnames]{color}
\usepackage{cancel}
\usepackage{soul} 
\usepackage[mathscr]{eucal}
\usepackage[T1]{fontenc}
\usepackage{braket}

\newcommand{\Real}{\mathbb{R}}
\newcommand{\Complex}{\mathbb{C}}

\newcommand{\A}{\mathcal{A}}
\newcommand{\C}{\mathcal{C}}
\newcommand{\D}{\mathcal{D}}
\newcommand{\Hop}{H}  
\newcommand{\Hilb}{\mathfrak{H}} 
\def\virg{\textquotedblleft}

\newcommand{\uu}{\mathscr{U}}
\newcommand{\vv}{\mathscr{V}}

\begin{document}

\title{Gauge-invariant spherical linear perturbations of wormholes in Einstein gravity minimally coupled to a self-interacting phantom scalar field}

\author{Francesco Cremona}
\affiliation{Dipartimento di Matematica, Universit\`a di Milano, Via C. Saldini 50, I-20133 Milano, Italy.}
\email{francesco.cremona@unimi.it}

\author{Livio Pizzocchero}
\affiliation{Dipartimento di Matematica, Universit\`a di Milano, Via C. Saldini 50, I-20133 Milano, Italy \\
and Istituto Nazionale di Fisica Nucleare, Sezione di Milano, Via G. Celoria 16, I-20133 Milano, Italy.}
\email{livio.pizzocchero@unimi.it}

\author{Olivier Sarbach}
\affiliation{Instituto de F\'\i sica y Matem\'aticas,
Universidad Michoacana de San Nicol\'as de Hidalgo,\\
Edificio C-3, Ciudad Universitaria, 58040 Morelia, Michoac\'an, M\'exico.}
\email{olivier.sarbach@umich.mx}

\begin{abstract}
Recently, there has been quite a lot of interest in static, spherical wormhole spacetimes and the question of their stability with respect to time-dependent perturbations. The consideration of linearized perturbations usually leads to a master wave equation with effective potential which can then be analyzed using standard tools from quantum mechanics. However, in the wormhole case, particular care must be taken with the gauge conditions when formulating the master equation. A poor coordinate choice, based for example on fixing the areal radial coordinate, may lead to singularities at the throat which complicate the stability analysis or might even lead to erroneous conclusions regarding the stability of the underlying wormhole configuration. In this work, we present a general method for deriving a gauge-invariant wave system of linearized perturbation equations in the spherically symmetric case, assuming that the matter supporting the wormhole is a phantom scalar field, that is, a self-interacting scalar field whose kinetic energy has the reversed sign. We show how this system can be decoupled and reduced to a single master wave equation with a regular potential, with no intermediate steps involving singularities at the throat. Two applications of our formalism are given. First, we rederive the master equation for the linearly perturbed Ellis-Bronnikov wormhole using our new, singularity-free method. As a second application, we derive the master equation describing the linear perturbations of a certain Anti de Sitter wormhole, provide a detailed analysis of the spectral properties of the underlying operator and prove that, as in the Ellis-Bronnikov case, this wormhole is linearly unstable and possesses a single unstable mode. In the final part of the paper, we consider a wormhole with de Sitter-type ends, whose spacetime presents horizons and admits a nonstatic extension beyond them; for this system we derive partial results of linear instability.
\end{abstract}

\maketitle

\section{Introduction}

One of the most fascinating features of Einstein's theory of General Relativity (GR) consists in the fact that spacetime may be curved and topologically nontrivial, describing intriguing objects like black holes and wormholes. Black hole spacetimes appear under rather natural conditions in GR, and they are expected to form in nature, for instance, by the collapse of sufficiently massive stars at the end of their life. Furthermore, there is by now compelling evidence for their existence in our Universe which has recently been reinforced by the observation of gravitational waves from binary black hole mergers~\cite{Abbott:2016blz} and the first image of the shadow of the supermassive black hole in the center of the galaxy M87~\cite{Akiyama:2019cqa}. In contrast to this, the occurrence of wormholes (\footnote{In this article, when talking about wormholes, we always refer to \emph{traversable} Lorentzian wormhole spacetimes in a metric theory of gravity.}) is much more speculative, and so far, there is no observational evidence for the existence of such structures. From the theoretical point of view, there are important constraints, such as the topological censorship theorem~\cite{jFkSdW93}. This theorem implies that asymptotically flat, globally hyperbolic wormhole spacetimes (including those relevant to this paper whose Cauchy surfaces have topology $\Real\times S^2$, representing a throat connecting two asymptotically flat ends) require the existence of ``exotic'' matter to support the throat, that is, they require matter whose stress-energy-momentum tensor violates the (averaged) null energy condition. Intuitively, the need for exotic matter can be understood by the fact that a light bundle that traverses a wormhole throat must focus as it approaches the throat, but then must expand again as it moves away from the throat, which is opposite to the focusing effect for light due to ordinary matter~\cite{HawkingEllis-Book}. On the other hand, it has also been shown that an infinitesimally small quantity of matter violating the averaged null condition is sufficient to support the throat~\cite{mVsKnD03}. This leads to the hope that quantum effects may give rise to a semiclassical theory in which wormhole spacetimes are allowed, in a similar way to how quantum effects (Hawking radiation) induce black hole evaporation although an area decrease of the event horizon is forbidden in classical GR with matter fields satisfying the null energy condition~\cite{HawkingEllis-Book}. Nevertheless, it remains to be seen whether or not such effects are strong enough to give rise to a traversable wormhole throat of macroscopic size~\cite{eFrW96}.

Instead of invoking quantum effects, an alternative way to violate the null energy condition (which has received important motivation from cosmology, see for example~\cite{Caldwell:1999ew}) is the consideration of phantom scalar fields, i.e. scalar fields that have a negative kinetic energy (see for instance~\cite{Bro2018} and references therein). Due to this property, such fields may lead to gravitational repulsion, and hence induce interesting effects like the accelerated expansion in the Universe, universes with no particle horizon~\cite{dFmGlP19} or the ability of supporting a wormhole throat~\cite{hE73,kB73}. On the other hand, the presence of unbounded negative kinetic energy might cast doubt on the possibility that any stationary solution found in this theory could ever be stable. (\footnote{However, the presence of unbounded negative kinetic energy by itself does not imply that any stationary solution in the theory is \emph{necessarily} unstable. For example, it turns out that the Minkowski spacetime is nonlinearly stable in Einstein theory minimally coupled to a massless scalar field irrespectively of the sign of the gravitational coupling constant (see the comments and references in appendix B.5 of~\cite{mDiR08}).}) Therefore, a pressing question regarding the relevance of static wormhole solutions in such theories (or other GR theories involving exotic matter fields) is their dynamical stability under small perturbations.

The most widely studied wormhole models (including those analyzed in the present article) are based on static, spherically symmetric spacetimes in which the world sheet of the throat consists of spheres of minimal area~\cite{mMkT88}. Within the context of phantom scalar fields, many such solutions have been found; the simplest ones are obtained for a real scalar field and are due to pioneering work by Ellis~\cite{hE73} and by Bronnikov~\cite{kB73}.
Since then, these solutions have been generalized to arbitrary dimensions~\cite{jEtZ08,Torii}
and to the following supporting fields: a scalar with a self-interaction potential~\cite{Dzhunushaliev:2008bq,kBrKaZ12}, a complex phantom scalar~\cite{Dzhunushaliev:2017syc}, a family of conventional and/or phantom scalars~\cite{hSsH02,oStZ12,Carvente:2019gkd}, a phantom scalar and an electromagnetic field~\cite{jGfGoS09c}, and, very recently, a k-essence scalar \cite{kBjFdR2019}. For the linear stability analysis of many of these solutions, see Refs.~\cite{jGfGoS09a,jGfGoS09c,oStZ12,kBjFaZ11,kBrKaZ12,fCfPlP19,kBjFdR2019}. All these studies conclude that the static, spherically symmetric wormhole solutions are linearly unstable, with numerical simulations~\cite{hSsH02,jGfGoS09b,aDnKdNiN09} revealing that the throat either collapses to a black hole or expands on timescales comparable to the light-crossing time of the radius of the throat. Therefore, finding a static, spherically symmetric wormhole solution in GR with exotic matter which can be shown to be linearly stable (or unstable with a large timescale associated with all the unstable modes) remains a challenging open problem. (\footnote{See also~\cite{pKbKjK11} for the construction of static, spherically symmetric wormholes in Einstein-dilaton-Gauss-Bonnet theory, a modified gravity theory, which does not require exotic matter. However, a careful stability analysis has recently revealed that these solutions are linearly unstable as well~\cite{mCrKaZ18}.})

In this work, we focus on GR minimally coupled to a single, real phantom scalar field $\Phi$ with an arbitrary self-interaction potential $V(\Phi)$ and provide a general, gauge-invariant framework to analyze the linear stability of static, spherically symmetric wormhole solutions in these theories; the latter is tested in specific applications.

In order to clarify which are the novelties of the paper, it is necessary to sketch
the previous state of the art in this area. Linearized perturbations of wormhole
solutions of the Einstein's equations have been previously discussed, even in a gauge-invariant language.
However, most of the previous approaches are based
on fixing the radial coordinate and deriving a linearized wave equation for perturbations of the scalar field;
due to the fact that the radial coordinate has a critical point at the throat,
the effective potential appearing in this wave equation is necessarily \emph{singular} at the throat.
As explained in~\cite{jGfGoS09a} (see also~\cite{kBjFaZ11}) this yields an artificial (mirrorlike)
boundary condition at the throat which prevents perturbations from traversing the wormhole.
This artificial boundary condition effectively restricts the class of physically admissible perturbations,
and as it turns out, the unstable modes associated with the wormhole are precluded from this class,
leading to the erroneous conclusion that the wormhole is linearly stable.
To overcome these problems, a method for transforming the singular wave equation
to a regular one was introduced in~\cite{jGfGoS09a} to treat the linearized perturbations
of the Ellis-Bronnikov wormhole; this approach was subsequently generalized
and referred to as ``S-deformation method'' in~\cite{kBjFaZ11}. Both \cite{jGfGoS09a}
and \cite{kBjFaZ11} refer to $(3+1)$- dimensional spacetimes; higher dimensional
extensions were considered in \cite{Torii}, where the
reflection-symmetric Ellis-Bronnikov wormhole solution was generalized
to any spacetime dimension $d+1$ (with $d \geqslant 3$) and its linear
stability analysis was performed, using again the S-deformation method
to overcome singularity problems at the throat and eventually showing that the wormhole
under consideration is unstable in any dimension.

We are now ready to describe the novelties of the present paper.
Here we work in spacetime dimension $3+1$, in
the framework already outlined
(a phantom scalar with self-interaction minimally coupled
to gravity, the spherically symmetric wormhole solutions arising from this
setting and their linear stability analysis). Our first result
is the derivation of a coupled, $2\times 2$ linear wave system subject to a constraint,
describing the linearized dynamics of time-dependent perturbations of such solutions in terms of two gauge-invariant
linear combinations of the linearized metric coefficients and of the scalar field;
a key feature of this system is that
it is \emph{regular} at the throat, provided the scalar field does not have a critical point there.
The second result of our work is that,
provided a nontrivial time-independent solution of the coupled $2\times 2$ system is known,
it is possible to decouple the system, obtaining a single wave equation for an appropriate,
gauge-invariant linear combination of the perturbed fields,
from which all other perturbations can be reconstructed;
in most situations, such a time-independent solution can be found by
varying the parameters of a known family of static wormhole
solutions. The above two results provide a general frame for spherically
symmetric wormholes and their linear stability analysis, alternative
to the S-deformation approach of \cite{jGfGoS09a} \cite{kBjFaZ11} \cite{Torii}:
no S-deformation of the linearized perturbation equations is necessary in the approach
of this paper, since there is no singularity to be eliminated.

For the Ellis-Bronnikov wormhole, we show that the master equation obtained by our method agrees precisely with the one obtained in~\cite{jGfGoS09a} by the S-method. Furthermore, we show that our gauge-invariant method for obtaining a master equation through the decoupling of the $2\times 2$ system also works for wormhole solutions whose stability has not been addressed so far. As an explicit example, we consider a static, spherically symmetric Anti de Sitter (AdS)-type wormhole which connects two asymptotic AdS ends (this is a special case of a family of static solutions of the Einstein-scalar equations derived by Bronnikov and Fabris in~\cite{kBjF06}\cite{Bro2018}); we prove that the above AdS wormhole is linearly unstable, a fact that we presume to be a third novelty of the present work.

Finally, in this paper we provide a detailed analysis for the behavior of the solution of the master equations in both the Ellis-Bronnikov and the AdS case, based on a rigorous spectral analysis of the Schr\"odinger operator appearing therein. A negative eigenvalue of the Schr\"odinger operator gives rise to a pair of modes, one exponentially growing and the other one exponentially decaying with respect to the time variable;
a positive eigenvalue gives rise to a pair of oscillating modes, while a positive energy level
lying in the continuous spectrum gives rise to a pair of non-normalizable oscillating modes,
corresponding to generalized eigenfunctions of the Schr\"odinger operator.
If zero is an eigenvalue it gives rise to a pair of normalizable modes,
one of them constant and the other one linearly growing with time.
We show that in the Ellis-Bronnikov case, the solution can be expanded in terms of an exponentially growing, an exponentially decaying, a constant, a linearly growing mode
and a continuum of oscillators associated with non-normalizable modes.
In contrast to this, in the AdS case the spectrum is a pure point spectrum giving rise to an exponentially growing, an exponentially decaying, and to an infinite, discrete set of oscillating normalizable modes. This is due to the Dirichlet-type boundary conditions imposed at the AdS boundary, which give rise to a regular Sturm-Liouville problem.

The AdS wormhole has a de Sitter (dS) analog which, however, presents horizons; to go beyond the horizons it is necessary to consider a Kruskal-type extension of the dS wormhole spacetime, which, however, is nonstatic and thus is outside the mainstream of the paper. In any case, in the final part of the paper we discuss the above issues and also present a partial result of linear instability, concerning the static part of the wormhole spacetime (we think this is another novelty of this article, foreshadowing future developments).

The article is organized as follows. In section~\ref{Sec:SphSymFEQ} we specify our metric ansatz, make a few general comments regarding the coordinate conditions that will be relevant in this work and derive the field equations for a spherically symmetric, time-dependent configuration. In section~\ref{Sec:Static} we mainly discuss two static wormhole solutions that will serve as examples and applications for our perturbation formalism and stability analysis: the Ellis-Bronnikov solution and the previously mentioned wormhole between two AdS universes. In the same section we spend a few words on the dS analog of this wormhole, to be reconsidered in the final part of the article. In section~\ref{Sec:LinearPerturbation} we derive the relevant set of linearized equations in a gauge-fixed setting in which the scalar field is held fixed. In section~\ref{Sec:GaugeInvariant} we introduce a set of combinations of the linearized fields which are invariant with respect to infinitesimal coordinate transformations, and the linearized field equations are cast into a constrained wave system for two of these gauge-invariant fields. In section~\ref{Sec:Decoupling} we show how to decouple this wave system, provided a static solution of the linearized field equations is available, in which case a single master wave equation is obtained. This method is then applied to the examples of  section~\ref{Sec:Static}, and it is shown that in each case the associated Schr\"odinger operator possesses a unique bound state with negative energy, implying that these wormholes are linearly unstable. In section~\ref{Sec:Spectral} we provide a detailed discussion on the spectral decomposition of the Schr\"odinger operator and the corresponding master equations (based on rigorous techniques from functional analysis) and contrast the Ellis-Bronnikov case with the one of the AdS wormhole. In section~\ref{Section:dS} we describe the dS wormhole, including the nonstatic extension beyond the horizons of its spacetime; we also derive a linear instability result concerning the static part of this spacetime. Conclusions, limitations and possible future applications of our method are given in section~\ref{Sec:Conclusions}. Technical details regarding the spectral theory of Schr\"odinger operators are given in the appendices.

Throughout this work, we use the signature convention $(-,+,+,+,)$ and choose units  in which $c=1$, $\hbar=1$.

\section{Spherically symmetric field equations and background}
\label{Sec:SphSymFEQ}

We consider a four-dimensional spacetime $(M,{\bf g})$ in which the gravitational field ${\bf g}$ is minimally coupled to a massless phantom scalar field $\Phi$, that is, a scalar field with the reversed sign in its kinetic term that self-interacts according to a potential $V(\Phi)$. The action functional of this system is
\[S[\mathbf{g},\Phi]:=\int\bigg(\frac{R}{2\kappa}
 +\frac{1}{2} \nabla^\mu\Phi\cdot \nabla_\mu \Phi -V(\Phi)\bigg)dv\,,  \]
where $\kappa = 8\pi G$ is the usual coupling constant while $R$ and
$dv=\sqrt{|\textnormal{det}(g_{\mu\nu})|}\prod\limits_{\mu=0}^3 dx^{\mu}$ are the scalar curvature and the volume element associated with the metric $\mathbf{g}$. The corresponding field equations are
\begin{eqnarray}
R_{\mu\nu} &=& \kappa\left[ -\nabla_\mu\Phi \cdot \nabla_\nu\Phi + V(\Phi) g_{\mu\nu} \right]\,,
\label{Eq:Einstein}\\
0 &=&  \nabla^\mu\nabla_\mu\Phi + V'(\Phi)\,,
\label{Eq:KleinGordon}
\end{eqnarray}
with $\nabla_\mu$ and $R_{\mu\nu}$ denoting the covariant
derivative and Ricci tensor, respectively, associated with ${\bf
g}$.

In this work, we focus on spherically symmetric spacetimes $(M,{\bf g})$ of the form $M = \tilde{M}\times S^2$ with metric
\begin{equation}
{\bf g} = -\alpha(t,x)^2 dt^2 + \gamma(t,x)^2 \left( dx + \beta(t,x) dt \right)^2
 + r(t,x)^2\left( d\vartheta^2 + \sin^2\vartheta\; d\varphi^2 \right)\,,
\label{Eq:SphericalMetric}
\end{equation}
which, in a general spherically symmetric coordinate system $(t,x,\vartheta,\varphi)$, is parametrized in terms of the four functions $\alpha,\beta,\gamma,r$ on the two-dimensional manifold $\tilde{M}$. Of course, the number of these functions can be reduced from four to two by an appropriate choice of the coordinates $(t,x)$ on $\tilde{M}$. There are several ``natural'' choices one can make. For example, given a smooth function $f: \tilde{M}\to \Real$ with the property that its gradient is everywhere spacelike, one can choose an orthogonal coordinate system $(t,x)$ on $\tilde{M}$ such that $x = f$ and $\beta = 0$. (Likewise, if the gradient of $f$ is everywhere timelike one can choose $(t,x)$ such that $\beta = 0$ and $t = f$.) In particular, if the gradient of the areal radius $r$ is everywhere spacelike one can choose $f = r$ and one is left with the two functions $\alpha$ and $\gamma$ on $\tilde{M}$. Usually, however, the gradient of $r$ is not everywhere spacelike due to the presence of minimal or trapped surfaces, and the resulting coordinate system is only locally defined on $\tilde{M}$.

The field equations~(\ref{Eq:Einstein},\ref{Eq:KleinGordon}) for a spherically symmetric metric~(\ref{Eq:SphericalMetric}) in any gauge such that $\beta = 0$ can be written as
\begin{eqnarray}
\frac{\partial}{\partial t}\left( \frac{\dot{\gamma}}{\alpha} \right)
 - \frac{\partial}{\partial x}\left( \frac{\alpha'}{\gamma} \right)
 - \frac{\gamma}{\alpha}\frac{\dot{r}^2}{r^2}
 + \frac{\alpha}{\gamma}\frac{r'^2}{r^2}
 - \frac{\alpha\gamma}{r^2}
 &=& \frac{\kappa}{2}\left[ \frac{\gamma}{\alpha}\dot{\Phi}^2 - \frac{\alpha}{\gamma}\Phi'^2 \right]\,,
\label{Eq:Ev1}\\
\frac{\partial}{\partial t} \left[ \frac{\gamma}{\alpha} r\dot{r} \right]
 - \frac{\partial}{\partial x} \left[ \frac{\alpha}{\gamma} r r' \right]
  &=& \alpha\gamma\left( \kappa r^2 V(\Phi)-1\right)\,,
\label{Eq:Ev2}\\
\frac{\partial}{\partial t} \left[ \frac{\gamma}{\alpha} r^2\dot{\Phi} \right]
 - \frac{\partial}{\partial x}\left[ \frac{\alpha}{\gamma} r^2\Phi' \right] &=& \alpha \gamma r^2 V'(\Phi)\,,
\label{Eq:Ev3}
\end{eqnarray}
with the constraints
\begin{eqnarray}
{\cal H} &:=& \frac{\alpha}{\gamma}\left[ 2\frac{r''}{r} + \frac{r'}{r}\left( \frac{r'}{r} - 2\frac{\gamma'}{\gamma} \right) \right]
 - \frac{\gamma}{\alpha}\frac{\dot{r}}{r}\left( \frac{\dot{r}}{r} + 2\frac{\dot{\gamma}}{\gamma} \right)
 - \frac{\alpha\gamma}{r^2}
 -\frac{\kappa}{2}\left[ \frac{\gamma}{\alpha}\dot{\Phi}^2 + \frac{\alpha}{\gamma}\Phi'^2 \right]
  + \kappa\alpha \gamma V(\Phi) = 0\,,
\label{Eq:Evh}
 \\
{\cal M} &:=& 2\frac{\dot{r}'}{r} - 2\frac{\dot{r}}{r}\frac{\alpha'}{\alpha}
 - 2\frac{r'}{r}\frac{\dot{\gamma}}{\gamma}
  -\kappa\dot{\Phi}\Phi' = 0\,.
\label{Eq:Evm}
\end{eqnarray}
Here and in the following, a dot and a prime refer to partial differentiation with respect to $t$ and $x$, respectively. In the conformally flat gauge, in which $\alpha=\gamma$, Eqs.~(\ref{Eq:Ev1},\ref{Eq:Ev2},\ref{Eq:Ev3}) yield a hyperbolic wave system for the quantities $(\alpha,r,\Phi)$ which is subject to the constraints~(\ref{Eq:Evh},\ref{Eq:Evm}). This system (or slight variants thereof) is suitable for numerical time evolutions, see for instance~\cite{jGfGoS09b}.

\section{Static wormhole solutions}
\label{Sec:Static}

In this section we deal with some examples of static wormhole solutions that have been considered previously in the literature. Most of our attention will be devoted to the Ellis-Bronnikov wormhole connecting two asymptotically flat ends~\cite{hE73,kB73} and to a reflection-symmetric wormhole connecting two AdS ends~\cite{Bro2018}; these will be the main applications of the general technique for linear stability analysis proposed in the present work (sections \ref{Sec:LinearPerturbation}-\ref{Sec:Spectral}). We will also mention a wormhole with ``dS-asymptotics'' \cite{Bro2018}; this case is essentially different from the previous two since it has horizons, a feature which is essentially outside the mainstream of the present work. We will return to this dS wormhole in section \ref{Section:dS}, where we will give a first draft of the treatment of this wormhole, including hints on its linear stability analysis; we hope to reconsider this subject in future works.

\subsection{Ellis-Bronnikov wormhole}
\label{Subsection:Ellis}

Let us assume a zero potential: $V(\Phi)=0$. In the static case, the functions $\alpha$, $\gamma$ and $r$ are $t$-independent and one
can further adjust the coordinate $x$ so that $\alpha\gamma = 1$. In this case, the field equations can be reduced to the three differential equations
$$
[ \alpha^2 r^2]'' = 2\,,\qquad
[ \alpha^2 r r']' = 1\,,\qquad
[\alpha^2 r^2\Phi']' = 0\,,
$$
which arise, respectively, from a recombination of Eqs.~(\ref{Eq:Ev1},\ref{Eq:Ev2},\ref{Eq:Evh}),
from Eq.~(\ref{Eq:Ev2}) and from Eq.~(\ref{Eq:Ev3}).
These can easily be integrated with the result
\begin{equation}
\alpha = \gamma^{-1} = e^{\gamma_1 \arctan {x \over b}}\,,\qquad
r^2 = (x^2 + b^2)\gamma^2\,,\qquad
\Phi = \Phi_1\arctan {x \over b}\,.
\label{Eq:StaticSolutions}
\end{equation}
Here, $b > 0$, $\gamma_1$ and $\Phi_1$ are integration constants, and the Hamiltonian constraint ${\cal H} = 0$ enforces the relation $\kappa\Phi_1^2 = 2(1 + \gamma_1^2)$ (while the momentum constraint ${\cal M}=0$ is obviously satisfied). This solution was obtained long time ago by Ellis~\cite{hE73} and Bronnikov~\cite{kB73} and describes a traversable wormhole whose throat is located at $x = \gamma_1 b$ (see also~\cite{jGfGoS09a} for its physical properties). The reflection-symmetric case $\gamma_1 = 0$ for which $\alpha = \gamma = 1$ results in a particularly simple form of the wormhole metric which has been posed as an exercise in general relativity in the popular article by Morris and Thorne~\cite{mMkT88}.

\subsection{A wormhole connecting two AdS universes}
\label{Subsection:AdS}

We now look for a static solution in the gauge $\alpha \gamma=1$, allowing $V(\Phi)$ to be nonzero. Let us show that a simple solution of this form can be obtained by setting as before
$r = \sqrt{x^2+b^2}$, where $b>0$. With these choices it is easy to show that the
combination (Eq.~(\ref{Eq:Ev2})$+r^2$Eq.~(\ref{Eq:Evh})) is satisfied if $\Phi = \sqrt{2/\kappa}\arctan(x/b)+\Phi_0$
with $\Phi_0$ a constant. With this expression for the scalar field, Eq.~(\ref{Eq:Ev1})
leads to
$$
\alpha=\sqrt{1-K \left(b^2+x^2\right)+M \left(b^2+x^2\right) \arctan\frac{x}{b} + b M x}\,,
$$
where $K$ and $M$ are two constants. The remaining two equations, Eqs.~(\ref{Eq:Ev2},\ref{Eq:Ev3}) (or, alternatively, Eqs.~(\ref{Eq:Ev3},\ref{Eq:Evh})), can be solved by setting
$$
V(\Phi(x))=\frac{ K(b^2+3 x^2)   -M \left(b^2+3 x^2\right) \arctan\frac{x}{b} - 3 b M x}{\kappa\left(b^2+x^2\right)}\,.
$$
Choosing, without loss of generality, $\Phi_0=0$, we obtain for $V(\Phi)$
$$
V(\Phi)=\frac{K}{\kappa}\left[ 3 -2  \cos^2\left( \sqrt{\frac{\kappa}{2}}\Phi\right) \right]
 - \frac{M}{\kappa}\left\{ 3 \sin \left( \sqrt{\frac{\kappa}{2}}\Phi\right) \cos \left( \sqrt{\frac{\kappa}{2}}\Phi\right)+   \sqrt{\frac{\kappa}{2}}\Phi\left[ 3 -2  \cos ^2\left( \sqrt{\frac{\kappa}{2}}\Phi\right)\right]\right\}\,.
$$
Actually, this solution is exactly the general solution given by Bronnikov and Fabris in~\cite{kBjF06} and reconsidered in the recent survey~\cite{Bro2018} (with some reparametrization of the involved constants).

From here to the end of the paper we make the choice
\begin{equation}
M = 0\,
\label{Eq:ConstantM}
\end{equation}
corresponding to a wormhole metric which is reflection-symmetric with respect to the throat;
hereafter and in most of the paper we also set
\begin{equation}
K\equiv - k^2\,, \qquad (k>0)\,.
\label{Eq:ConstantK}
\end{equation}
With the choices (\ref{Eq:ConstantM}.\ref{Eq:ConstantK}), the solution simplifies to
\begin{equation}
V(\Phi)= -\frac{k^2}{\kappa}
\left[ 3 -2  \cos ^2\left( \sqrt{\frac{\kappa}{2}}\Phi\right) \right]\,,\quad
\alpha = \gamma^{-1} = \sqrt{1 + k^2(x^2+b^2)}\,,\quad
r = \sqrt{x^2 + b^2}\,,\quad
\Phi = \sqrt{\frac{2}{\kappa}}\arctan\frac{x}{b}\,.
\label{Eq:StaticSolutionsPot1}
\end{equation}
In the limit case $b\to0$ we should replace the third equality in~(\ref{Eq:StaticSolutionsPot1}) with $r=x>0$; the corresponding metric describes an AdS universe with cosmological constant $\Lambda = -3k^2$. From now on, we intend ($b>0$, as already stated and)
\begin{equation}
x\in(-\infty , +\infty)\,;
\end{equation}
since $r(x)\sim |x|$ for $x\to \pm \infty$, we can interpret the metric in~(\ref{Eq:StaticSolutionsPot1}) as describing a wormhole connecting two separate asymptotically AdS universes with the same cosmological constant $\Lambda = -3k^2$ and minimal areal radius $b$ at the throat.
For this reason one could call the solution~(\ref{Eq:StaticSolutionsPot1}) an ``AdS-AdS wormhole'';
in the sequel this expression will always be shortened to ``AdS wormhole''.
Let us note that, for $k\to0$, the potential $V(\Phi)$ vanishes and the AdS wormhole  (with $b$ fixed) becomes the reflection-symmetric Ellis-Bronnikov wormhole (as in Eq. (\ref{Eq:StaticSolutions}), with $\gamma_1=0$).

For further convenience, we introduce the change of variables
\begin{equation}
t = \frac{s}{2k \sqrt{1+B^2}}\,,\qquad x=\frac{\sqrt{1+B^2}}{k}\tan{\frac{u}{2}}\,,\qquad
B:=b k  \,,\qquad
s\in (-\infty,+\infty)\,,\quad u\in (-\pi,\pi)\,,
\label{coordsu}
\end{equation}
so that in the new coordinate system the metric corresponding to the solution~(\ref{Eq:StaticSolutionsPot1}) is transformed into a metric of the form~(\ref{Eq:SphericalMetric}) with $(t,x)$ replaced by $(s,u)$ and
\begin{equation}
\alpha=\gamma = \frac{1}{2 k \cos{\frac{u}{2}}}\,,\qquad \beta = 0\,,\qquad
r= \frac{\sqrt{1+2 B^2-\cos u}}{\sqrt{2} k \cos \frac{u}{2}}\,,\qquad\Phi=\sqrt{\frac{2}{\kappa}}\arctan\left(\frac{\sqrt{1+B^2}\tan \frac{u}{2}}{B} \right)
\label{Eq:StaticSolutionsPot2}
\end{equation}
(of course, $V(\Phi)$ is still as in~(\ref{Eq:StaticSolutionsPot1})). Let us observe that the limits $x\to\pm\infty$, describing the far ends of the wormhole, are equivalent to the limits $u\to\pm \pi$.

\subsection{A dS wormhole}
\label{Subsection:dS}

As anticipated in the first paragraph of this section, we will consider later a dS-type wormhole, differing substantially from the Ellis-Bronnikov and AdS wormholes due to the presence of horizons. For the moment, we just mention that this dS wormhole is the case $M=0$, $K= k^2>0$ of the Bronnikov-Fabris solution described at the beginning of section~\ref{Subsection:AdS}; we will return to this wormhole in section \ref{Section:dS}, after acquiring experience on linearized perturbation theory through the analysis of the AdS case.

\section{Linear perturbations and the $\delta\Phi = 0$ gauge}
\label{Sec:LinearPerturbation}

In the sequel we consider, for an arbitrary potential $V(\Phi)$, a family of static solutions $(\alpha,\gamma,r,\Phi)$ of Eqs.~(\ref{Eq:Ev1}-\ref{Eq:Evm}) (without necessarily assuming the gauge condition $\alpha\gamma = 1$). This family may depend on certain parameters (like the constants $b,\gamma_1$ in subsection~\ref{Subsection:Ellis} or the parameter $B$ in subsection~\ref{Subsection:AdS}). In addition, we consider a (nonstatic) perturbation $(\delta\alpha,\delta\gamma,\delta r,\delta\Phi)$ of this static solution, which is treated by linearizing Eqs.~(\ref{Eq:Ev1}-\ref{Eq:Evm}); let us recall that Eqs.~(\ref{Eq:Ev1}-\ref{Eq:Evm}) assume $\beta = 0$ for the metric~(\ref{Eq:SphericalMetric}), so their linearization corresponds to taking $\delta \beta=0$.

For the particular case in which the potential vanishes ($V = 0$) it can be shown (see e.g. Ref.~\cite{jGfGoS09a}) that the linearized constraint equations $\delta {\cal H} = \delta {\cal M} = 0$ can be integrated. It turns out this is still the case for solutions with a nontrivial potential, yielding the conclusion that
\begin{equation}
\sigma:={\alpha r \over \gamma}\left(
\delta r'-\frac{\alpha'}{\alpha}\delta r
-r'\frac{\delta\gamma}{\gamma}- \frac{\kappa}{2} r\Phi'\delta\Phi
\right)
\label{Eq:Pert1}
\end{equation}
is a constant. This constant indeed describes a zero mode, that is,
a perturbation corresponding to an infinitesimal variation of
the parameters labeling the static solution (see section 3.1 of~\cite{jGfGoS09a}
for more details in the $V=0$ case). Since we are mainly interested in dynamical perturbations (rather than infinitesimal deformations along the static branch in the solution space), we assume from now on that
\begin{equation}
\sigma = 0\,.
\label{Eq:Sigma0}
\end{equation}
For future use, it is advantageous to introduce the quantities (\footnote{This choice of notation is somehow awkward; however the reason for it is to maintain compatibility with the notation used in Ref.~\cite{jGfGoS08}.})
\begin{equation}
\D := \frac{\delta\alpha}{\alpha}\,,\qquad
\A := \frac{\delta\gamma}{\gamma}\,,\qquad
\C := \frac{\delta r}{r}\,.
\label{Eq:DAC}
\end{equation}
Then Eqs.~(\ref{Eq:Pert1}-\ref{Eq:Sigma0}) become
\begin{equation}
\sigma=0\,,\qquad \sigma:={\alpha r^2 \over \gamma}\left[ \C' - \left(\frac{\alpha'}{\alpha} - \frac{r'}{r} \right) \C - \frac{r'}{r} \A - \frac{\kappa}{2}\Phi'\delta\Phi\right]\,;
\label{Eq:Pert1Bis}
\end{equation}
moreover, the linearization of Eqs.~(\ref{Eq:Ev1},\ref{Eq:Ev2},\ref{Eq:Ev3}) and the condition $\sigma=0$ give the following linear system of equations:
\begin{eqnarray}
 \frac{\gamma}{\alpha}\ddot{\A} - \frac{\partial}{\partial x}\left( \frac{\alpha}{\gamma} \D' \right)
 - \frac{\alpha'}{\gamma}(\D - \A)' + 2\frac{\alpha}{\gamma}\frac{r'}{r} \C'
  - \frac{2\alpha\gamma}{r^2}(\A - \C) + \kappa\frac{\alpha}{\gamma}\Phi'\delta\Phi' &=& 0\,,
\label{Eq:A}\\
 \frac{\gamma}{\alpha}\ddot{\C} - \frac{\partial}{\partial x}\left( \frac{\alpha}{\gamma} \C' \right)
 - \frac{\alpha}{\gamma}\frac{r'}{r}(\D - \A + 4\C)' + \frac{2\alpha\gamma}{r^2}(\A - \C)
  -\kappa\alpha\gamma\left[2V(\Phi) \A + V'(\Phi)\delta\Phi \right] &=& 0\,,
\label{Eq:C}\\
  \frac{\gamma}{\alpha}\ddot{\delta\Phi}
 - \frac{\partial}{\partial x}\left( \frac{\alpha}{\gamma} \delta\Phi' \right)
 - 2\frac{\alpha}{\gamma}\frac{r'}{r}\delta\Phi' - \frac{\alpha}{\gamma}\Phi'( \D - \A + 2\C)'
  - \alpha\gamma\left[ 2V'(\Phi) \A + V''(\Phi)\delta\Phi \right] &=& 0\,.
\label{Eq:Phi}
\end{eqnarray}
 All the equations derived so far only assume the orthogonal gauge $\beta = 0$; at the linearized level there is still liberty which is related to the choice of a function $f$ on $\tilde{M}$, as explained in section~\ref{Sec:SphSymFEQ}. One possible choice is fixing the areal radius function $r(x)$ to its background form, such that $\delta r = 0$ and $\C = 0$. Equations~(\ref{Eq:Pert1Bis},\ref{Eq:C})  then allow us to express the metric fields $\A$ and $\D'$ in terms of $\delta\Phi$, and we obtain a master equation for the linearized scalar field $\delta\Phi$ (see e.g.~\cite{Bro2018}).

However, in this article, we are interested in deriving a master equation describing the dynamics of the linear perturbations of any one of the two wormholes in the previous subsections~\ref{Subsection:Ellis} and \ref{Subsection:AdS}. Since these solutions have $dr = 0$ at the wormhole throat, fixing the areal radius function $r(x)$ amounts to forcing the perturbations to vanish at the throat, which from a physical point of view is much too restrictive. At the mathematical level, enforcing the $\delta r = 0$ gauge results in a master equation for $\delta\Phi$ with a potential that is singular at the throat (see~\cite{jGfGoS09a,Bro2018} for more details). On the other hand, while $dr = 0$ at the wormhole throat, we note from Eq.~(\ref{Eq:StaticSolutions}) or Eq.~(\ref{Eq:StaticSolutionsPot1}) that $d\Phi =$ const$ \times dx/(x^2 + b^2)$ is everywhere spacelike, and hence the same will be true for sufficiently small perturbations of the static wormhole solution. As a consequence, we may choose the coordinates $(t,x)$ such that $\Phi$ is given by exactly the same expression as in Eq.~(\ref{Eq:StaticSolutions}) or Eq.~(\ref{Eq:StaticSolutionsPot1}), even for the perturbed spacetime. This implies, in particular, that $\delta\Phi = 0$. In this gauge, Eqs.~(\ref{Eq:Pert1Bis},\ref{Eq:Phi}) reduce to
\begin{eqnarray}
&& \sigma=0,\qquad \sigma:={\alpha r^2 \over \gamma}\left[ \C' - \left(\frac{\alpha'}{\alpha} - \frac{r'}{r} \right) \C - \frac{r'}{r} \A \right]\,,
\label{Eq:Pert3a}\\
&& \D' - \A' + 2\C' + 2 \gamma^2{ V'(\Phi)\over \Phi'} \A = 0\,;
\label{Eq:Pert3b}
\end{eqnarray}
using these equations in order to eliminate $\C'$ and $\D'$ and the static version of Eq.~(\ref{Eq:Ev3}) (from which one can eliminate the unperturbed quantity $\Phi''$), Eqs.~(\ref{Eq:A},\ref{Eq:C}) reduce to
\begin{eqnarray}
 \frac{\gamma}{\alpha}\ddot{\A}
 - \frac{\partial}{\partial x}\left[ \frac{\alpha}{\gamma}(\A' - 2\C') \right]
 + 2\frac{\alpha}{\gamma}\left( \frac{\alpha'}{\alpha} + \frac{r'}{r} \right) \C'
  {- 2\frac{\alpha\gamma}{r^2}(\A - \C)}
\nonumber\\
  \qquad\qquad + {2\alpha\gamma V'(\Phi) \over  \Phi'} \A'
  + 2\alpha\gamma\left[ \gamma^2{V'(\Phi)^2\over\Phi'^2}
  + \left( 3\frac{\alpha'}{\alpha} + 2\frac{r'}{r} \right) {V'(\Phi) \over \Phi'} + V''(\Phi) \right] \A &=& 0\,,
\label{Eq:Pert4}\\
 \frac{\gamma}{\alpha}\ddot{\C}
 - \frac{\partial}{\partial x}\left[ \frac{\alpha}{\gamma} \C' \right]
 - 2\frac{\alpha}{\gamma}\frac{r'}{r} \C'
 +2\frac{\alpha\gamma}{r^2}(\A - \C)
 + 2\alpha\gamma \left[ \frac{r'}{r} {V'(\Phi)\over \Phi'} - \kappa V(\Phi) \right] \A &=& 0\,,
\label{Eq:Pert5}
\end{eqnarray}
which is still subject to the constraint given in Eq.~(\ref{Eq:Pert3a}).

As a simple example, let us consider the reflection-symmetric subcase of the Ellis-Bronnikov wormhole, already mentioned at the end of the subsection~\ref{Subsection:Ellis}, corresponding to the choice $V = 0$, $\alpha=\gamma=1$ and $r$ as in Eq.~(\ref{Eq:StaticSolutions}). In this case, the difference between Eqs.~(\ref{Eq:Pert4},\ref{Eq:Pert5}), along with Eq.~(\ref{Eq:Pert3a}), gives
\begin{equation}
\label{Eq:MasterEllis}
\ddot{\chi} - \chi'' - \frac{3b^2}{r^4}\chi = 0\,,\qquad \chi := \frac{\A - \C}{r}\,,
\end{equation}
which coincides with Eq.~(15) in Ref.~\cite{jGfGoS08}. We will return to this subcase at the end of section~\ref{Sec:GaugeInvariant}. The generalizations to the non-reflection-symmetric Ellis-Bronnikov wormhole and to the AdS wormhole (subsections~\ref{Subsection:Ellis},  \ref{Subsection:AdS}) will be discussed in section~\ref{Sec:Decoupling}.

\section{Gauge-invariant reinterpretation}
\label{Sec:GaugeInvariant}

In this section we analyze the behavior of the perturbed fields under an infinitesimal coordinate transformation
\begin{equation}
x^a \mapsto x^a + \delta x^a\,,\qquad (x^a) = (t,x)\,,
\label{Eq:InfCT}
\end{equation}
parametrized by a vector field $\mathbf{\delta x} = \delta x^a \partial_a = (\delta t) \partial_t + (\delta x) \partial_x$ on $\tilde{M}$,
and try to rewrite the equations from the previous section in terms of fields which
are manifestly gauge-invariant with respect to these transformations.

Under the transformation~(\ref{Eq:InfCT}) the linear perturbations of the radial part of the metric, $\tilde{g}_{ab} dx^a dx^b := -\alpha^2 dt^2 + \gamma^2(dx + \beta dt)^2$, of the areal radius $r$, and of scalar field $\Phi$ transform according to
$$
\delta \tilde{g}_{ab} \mapsto \delta \tilde{g}_{ab}  + \pounds_{\mathbf{\delta x}}\tilde{g}_{ab}\,,\qquad
\delta r \mapsto \delta r + \pounds_{\mathbf{\delta x}} r\,,\qquad
\delta\Phi \mapsto \delta\Phi + \pounds_{\mathbf{\delta x}}\Phi\,,
$$
with $\pounds_{\mathbf{\delta x}}$ denoting the Lie derivative with respect to $\mathbf{\delta x}$.
Parametrizing the metric as in Eq.~(\ref{Eq:SphericalMetric})
and assuming that the background is static, this yields
\begin{eqnarray}
\delta\alpha &\mapsto& \delta\alpha + \alpha'\delta x
 + \alpha \, \delta \dot{t}\,,\\
\delta\beta &\mapsto& \delta\beta + \delta \dot{x}
 - \frac{\alpha^2}{\gamma^2} \delta t'\,,\\
\delta\gamma &\mapsto&  \delta\gamma + \left( \gamma\delta x \right)'\,,\\
\delta r &\mapsto& \delta r + r'\delta x\,,\\
\delta\Phi &\mapsto& \delta\Phi + \Phi'\delta x
\end{eqnarray}
(where $\delta \dot{t}$, $\delta \dot{x}$ and $\delta t'$ refer to $\frac{\partial}{\partial t}(\delta t)$, $\frac{\partial}{\partial t}(\delta x)$ and $\frac{\partial}{\partial x}(\delta t)$, respectively; similar notations are used hereafter in relation to $\delta \beta$ and $\delta \Phi$).
The following three quantities are invariant with respect to these transformations:
\begin{eqnarray}
A &:=& \frac{\delta\gamma}{\gamma} - \frac{1}{\gamma}\left( \gamma\frac{\delta\Phi}{\Phi'} \right)'\,,
\label{Eq:AGI}\\
C &:=& \frac{\delta r}{r} - \frac{r'}{r}\frac{\delta\Phi}{\Phi'}\,,
\label{Eq:CGI}\\
E &:=& \left( \frac{\delta\alpha}{\alpha} \right)' - \left( \frac{\alpha'}{\alpha}\frac{\delta\Phi}{\Phi'} \right)'
 + \frac{\gamma^2}{\alpha^2} \left( \delta\dot{\beta} - \frac{\delta\ddot{\Phi}}{\Phi'} \right)\,.
\label{Eq:EGI}
\end{eqnarray}
In the particular gauge used in the second half of the previous section,
for which $\delta\beta = \delta\Phi = 0$, it turns out that $A=\A$, $C=\C$ and $E=\D'$, where $\A$, $\C$ and $\D$ are defined by Eq.~(\ref{Eq:DAC}). Therefore, in this gauge, we may replace the quantities $\A$, $\C$ and $\D'$ in Eqs.~(\ref{Eq:Pert3a}--\ref{Eq:Pert5}) with the quantities $A,C,E$ of the present section.
Since the linearized field equations are gauge-invariant, the equations obtained in this way are valid in \emph{any gauge}.\\ Summing up, our gauge-invariant equations are
\begin{eqnarray}
&& \sigma = 0,\qquad
\sigma := \frac{\alpha r^2}{\gamma} \left[
C'+\left({r'\over r}-{\alpha'\over\alpha}\right)C-{r'\over r}A
\right]\,,
\label{Eq:Pert3abis}\\
&&
E - A' + 2C'  + 2\gamma^2{V'(\Phi)\over \Phi'} A = 0\,,
\label{Eq:Pert3bbis}\\
&& \frac{\gamma}{\alpha}\ddot{A}
- \frac{\partial}{\partial x}\left[ \frac{\alpha}{\gamma}(A' - 2C') \right]
+ 2\frac{\alpha}{\gamma}\left( \frac{\alpha'}{\alpha} + \frac{r'}{r} \right) C'
{- 2\frac{\alpha\gamma}{r^2}(A - C)}
\nonumber\\
&& \qquad\qquad + {2\alpha\gamma V'(\Phi) \over  \Phi'} A'
+ 2\alpha\gamma\left[ \gamma^2{V'(\Phi)^2\over\Phi'^2}
+ \left( 3\frac{\alpha'}{\alpha} + 2\frac{r'}{r} \right) {V'(\Phi) \over \Phi'} + V''(\Phi) \right] A = 0\,,
\label{Eq:Pert4bis}\\
&& \frac{\gamma}{\alpha}\ddot{C}
- \frac{\partial}{\partial x}\left[ \frac{\alpha}{\gamma} C' \right]
- 2\frac{\alpha}{\gamma}\frac{r'}{r} C'
+2\frac{\alpha\gamma}{r^2}(A - C)
+ 2\alpha\gamma \left[ \frac{r'}{r} {V'(\Phi)\over \Phi'} - \kappa V(\Phi) \right] A = 0\,.
\label{Eq:Pert5bis}
\end{eqnarray}

As a simple example, consider again the reflection-symmetric Ellis-Bronnikov wormhole, for which $V = 0$, $\alpha=\gamma=1$ and $r = \sqrt{x^2 + b^2}$, i.e. the same example as the one described at the end of section~\ref{Sec:LinearPerturbation} with Eqs.~(\ref{Eq:Pert3a}-\ref{Eq:Pert5}) yielding Eq.~(\ref{Eq:MasterEllis}). However, now Eq.~(\ref{Eq:MasterEllis}) can be reinterpreted in a gauge-invariant framework where $\chi=(A-C)/r$. The interest of this equation is that it involves only one unknown function $\chi(t,x)$ and reduces the linear stability analysis of this wormhole to the spectral analysis of the Schr\"odinger operator $-d^2/dx^2-3b^2/(x^2 + b^2)^2$. Since this has one negative eigenvalue (see Refs. \cite{jGfGoS08,jGfGoS09a}), one concludes that the wormhole is unstable.

In Ref.~\cite{jGfGoS08} an attempt was made to provide the present gauge-invariant formulation of the field equations in this particular subcase: while the two gauge-invariant quantities $A$ and $C$ were correctly defined, the quantity $D = \delta\alpha/\alpha$ defined in~\cite{jGfGoS08} is only invariant with respect to the restricted set of gauge transformations for which $\delta \dot{t} = 0$. However, in general, this restricted set is not sufficient to achieve both conditions $\delta\Phi = 0$ and $\delta\beta = 0$ simultaneously,
on which the derivation in Ref.~\cite{jGfGoS08} was based.

Finally, let us observe that the the gauge-invariant Eqs.~(\ref{Eq:Pert3abis},\ref{Eq:Pert4bis},\ref{Eq:Pert5bis})
(again in the present subcase  $V = 0$, $\alpha=\gamma=1$ and $r = \sqrt{x^2 + b^2}$)
are related to the results presented in~\cite{fCfPlP19}, which are based on the gauge $\delta\alpha = \delta\beta = 0$. For example, Eq.~(\ref{Eq:Pert3abis}), when choosing the gauge $\delta\alpha = \delta\beta = 0$, yields Eq.~(3.9) in~\cite{fCfPlP19}. The analysis of~\cite{fCfPlP19} also yields a final equation similar to Eq.~(\ref{Eq:MasterEllis}), even though it uses a different approach related to the chosen gauge. (\footnote{In~\cite{fCfPlP19}, the variables $x$ and $b$ of the present paper are denoted with $\ell$ and $a$;
the field $\mathscr{R}$ fulfilling the master equation~(3.15) of the cited work is related
to the present gauge-invariant quantities $C$ and $E$ by the relation
$ \frac{\partial^2}{\partial t^2}\left[\mathscr{R}\left(\frac{t}{b},\frac{x}{b}\right)\right]
=\frac{r^2}{b^3}\left(r \ddot{C} - r' E\right)$. Using the linearized field equations, this can also be rewritten as $\frac{\partial^2}{\partial t^2}\left[\mathscr{R}\left(\frac{t}{b},\frac{x}{b}\right)\right] = -\frac{1}{br}(A - C)$, which explains why $\mathscr{R}$ satisfies the same master equation as $\chi$, up to a source term whose second time derivative vanishes.}  )

\section{Decoupling of the pulsation equations}
\label{Sec:Decoupling}

After considering the simple example of the reflection-symmetric Ellis-Bronnikov wormhole, let us return to the case of an arbitrary potential $V(\Phi)$. In this section we try to reduce the gauge-invariant equations of section~\ref{Sec:GaugeInvariant} to one involving only one unknown function $\chi(t,x)$ (generalizing the considerations which lead to Eq.~(\ref{Eq:MasterEllis}) in the reflection-symmetric Ellis-Bronnikov subcase). To this purpose we note the following: setting
\begin{equation}
{\cal F} := {A-C\over r}\,,\qquad{\cal G} := {C\over r}
\label{Eq:FandG}
\end{equation}
and performing a lengthy calculation, we can reformulate the system of Eqs.~(\ref{Eq:Pert3abis},\ref{Eq:Pert4bis},\ref{Eq:Pert5bis}) as the hyperbolic system of wave equations
\begin{equation}
\left[ \frac{\partial^2}{\partial t^2}
  - \left( \frac{\alpha}{\gamma}\frac{\partial}{\partial x} \right)^2
  +
  \left( \begin{array}{cc} Y_0 & Y_0 \\ 0 & 0 \end{array} \right)\frac{\alpha}{\gamma}\frac{\partial}{\partial x}
  + \frac{\alpha^2}{\gamma^2} \left( \begin{array}{cc} W_{11} & W_{12} \\ W_{21} & W_{22} \end{array} \right)
\right] \left( \begin{array}{cc} {\cal F} \\ {\cal G} \end{array} \right) = 0\,,
\label{Eq:WaveSystem}
\end{equation}
subject to the constraint
\begin{equation}
{\cal G}' = \left( \frac{\alpha'}{\alpha} - \frac{r'}{r} \right){\cal G} + \frac{r'}{r}{\cal F}\, .
\label{Eq:WaveConstraint}
\end{equation}
Here, the functions $ Y_0$ and $W_{ij}$ are given by the following functions of the background quantities:
\begin{eqnarray}
Y_0 &:=&2\alpha\gamma {V'(\Phi)\over{\Phi'}}\,,
 \label{Eq:Y0}\\
W_{11} &:=& \frac{r'}{r}\left( 4\frac{\alpha'}{\alpha} + 3\frac{r'}{r} \right) - 3\frac{\gamma^2}{r^2}
 + Z_{11}\,,\\
W_{12} &:=& 4\frac{\alpha'^2}{\alpha^2} +Z_{12}\,,\\
W_{21} &:=& -4\frac{r'^2}{r^2} + 2\frac{\gamma^2}{r^2} + Z_{21}\,,\\
W_{22} &:=& \frac{r'}{r}\left( -4\frac{\alpha'}{\alpha} + 3\frac{r'}{r} \right) - \frac{\gamma^2}{r^2} + Z_{22}\,,
\end{eqnarray}
where
\begin{eqnarray}
Z_{11} &:=& Z_{12} + \kappa\gamma^2V(\Phi)\,,\\
Z_{12} &:=& 2\gamma^2\left[ \gamma^2{V'(\Phi)^2\over \Phi'^2}
 + \left( 3{\alpha'\over \alpha}+2{r'\over r} \right){V'(\Phi) \over \Phi'} + V''(\Phi)  \right]\, ,\\
Z_{21} &:=& 2\gamma^2\left[ -\kappa V(\Phi) + {r'\over r}{V'(\Phi)\over \Phi'} \right]\,,\\
Z_{22} &:=& Z_{21} + \kappa\gamma^2 V(\Phi)\,.
\end{eqnarray}
In deriving the wave system~(\ref{Eq:WaveSystem}) we have used the background equations
\begin{eqnarray}
\frac{r'}{r}\left( 2\frac{\alpha'}{\alpha} + \frac{r'}{r} \right) - \frac{\gamma^2}{r^2}
 + \frac{\kappa}{2}\Phi'^2 + \kappa\gamma^2 V(\Phi) &=& 0\,,
 \label{Eq:Background1}\\
\frac{\alpha''}{\alpha}- \frac{\alpha'}{\alpha}\left( \frac{\gamma'}{\gamma} - 2\frac{r'}{r} \right)
 + \kappa \gamma^2 V(\Phi) & = & 0\,,
  \label{Eq:Background2}\\
\frac{r''}{r} - \frac{r'}{r}\left( \frac{\gamma'}{\gamma} - \frac{\alpha'}{\alpha}- \frac{r'}{r} \right) - \frac{\gamma^2}{r^2} + \kappa \gamma^2 V(\Phi) & = & 0\,.
 \label{Eq:Background3}
\end{eqnarray}
Observe that in the reflection-symmetric Ellis-Bronnikov subcase $V = 0$, $\alpha=\gamma=1$ and $r = \sqrt{x^2 + b^2}$ it follows that $Y_ 0 = W_{12} = 0$, such that the equation for ${\cal F}$ in the system~(\ref{Eq:WaveSystem}) decouples trivially from the remaining ones.

In the following, we describe a general trick which allows one to decouple the constrained wave system~(\ref{Eq:WaveSystem},\ref{Eq:WaveConstraint}). Let us suppose that we know a static solution $({\cal F}_0(x),{\cal G}_0(x))$ of Eqs.~(\ref{Eq:WaveSystem},\ref{Eq:WaveConstraint}) such that ${\cal G}_0(x)\neq 0$ for all $x$. In this case, based on Leibnitz's product rule,
it is not difficult to verify that the field
$$
\tilde\chi := {\cal F} - \frac{{\cal F}_0}{{\cal G}_0}{\cal G}
$$
satisfies the decoupled wave equation
\begin{equation}
\left[ \frac{\partial^2}{\partial t^2}
- \left( \frac{\alpha}{\gamma}\frac{\partial}{\partial x} \right)^2
+ Y_0\frac{\alpha}{\gamma} \frac{\partial}{\partial x}
 + \frac{\alpha^2}{\gamma^2}\tilde{\cal {V}} \right] \tilde\chi = 0\,,
\label{Eq:Master1}
\end{equation}
with the potential
\begin{equation}
\tilde{\mathcal{ V}} = W_{11} - \frac{{\cal F}_0}{{\cal G}_0} W_{21} - 2\frac{r'}{r}\left( \frac{{\cal F}_0}{{\cal G}_0} \right)'
  +\frac{\gamma}{\alpha}\frac{r'}{r} Y_0\left(   \frac{{\cal F}_0}{{\cal G}_0}+1  \right)\,.
\end{equation}

Let us observe that it is possible to eliminate the first spatial derivative in Eq.~(\ref{Eq:Master1}). Indeed, let us define
\begin{equation}
\label{Eq:ChiTrans}
\chi:={\tilde\chi\over a}\,,\qquad
a(x) = a_0 e^{\int_{x_0}^{x}\frac{Y_0(y)\gamma(y)}{2\alpha(y)}dy }\,,
\end{equation}
where $a_0$ and $x_0$ are two constants. (\footnote{Note that $a$ satisfies
$$
a' =  \frac{Y_0 \gamma}{2 \alpha} a\,,\qquad
a'' = \left[ \left(\frac{Y_0 \gamma}{2 \alpha}\right)'  + \left(\frac{Y_0 \gamma}{2 \alpha}\right)^2 \right] a\,.
$$
}) Then, it is found that $\chi$ satisfies the wave equation
\begin{equation}
\label{Eq:Master2}
\left[ \frac{\partial^2}{\partial t^2}
   - \left( \frac{\alpha}{\gamma}\frac{\partial}{\partial x} \right)^2
 + \frac{\alpha^2}{\gamma^2}{\cal {V}} \right] \chi = 0\,,
\end{equation}
with the potential (\footnote{Here $Y_0'$ can be computed by taking a derivative of Eq.~(\ref{Eq:Y0}) and eliminating $\Phi''$ via the static version of Eq.~(\ref{Eq:Ev3}). })
\begin{eqnarray}
{\cal{V}} &:=& \tilde{\mathcal{ V}}
 + \frac{1}{4}\frac{\gamma^2}{\alpha^2} Y_0^2 - \frac{1}{2}\frac{\gamma}{\alpha} Y_0'\nonumber\\
 &=& \frac{r'}{r}\left( 4\frac{\alpha'}{\alpha} + 3\frac{r'}{r} \right) -
  \frac{3\gamma^2}{r^2}
   + \gamma^2\left[ 2\gamma^2{V'(\Phi)^2\over \Phi'^2}
   + 4\left( \frac{\alpha'}{\alpha} + \frac{r'}{r} \right) {V'(\Phi) \over \Phi'}
   + V''(\Phi) + \kappa V(\Phi) \right]
 \nonumber\\
 &+& \left[ 4\frac{r'^2}{r^2} - \frac{2\gamma^2}{r^2} + 2\kappa\gamma^2 V(\Phi) \right]
 \frac{{\cal F}_0}{{\cal G}_0}
  - 2\frac{r'}{r}\left( \frac{{\cal F}_0}{{\cal G}_0} \right)'\,.
\label{Eq:Potential}
\end{eqnarray}
We refer to Eq.~(\ref{Eq:Master2}) as the \emph{master equation}; this reduces the linear stability analysis to the spectral analysis of the linear, Schr\"odinger-type operator $-\left( \frac{\alpha}{\gamma}\frac{d}{d x} \right)^2 + \frac{\alpha^2}{\gamma^2}\,{\cal {V}}$.

Once the master equation has been solved for the field $\chi(t,x)$, it is possible to reconstruct the gauge-invariant quantities ${\cal F}$ and ${\cal G}$ by integrating the constraint equation~(\ref{Eq:WaveConstraint}). Using the definition of $\chi$ and the fact that $({\cal F}_0,{\cal G}_0)$ satisfy the constraint, one obtains
\begin{eqnarray}
{\cal G}(t,x) &=& {\cal G}_0(x)\int\limits_{x_0}^x \frac{r'(y)}{r(y)}\frac{a(y)}{{\cal G}_0(y)} \chi(t,y) dy\,,\\
{\cal F}(t,x) &=& a(x)\chi(t,x) + \frac{{\cal F}_0(x)}{{\cal G}_0(x)} {\cal G}(t,x)\,,
\end{eqnarray}
and from this one can also reconstruct the gauge-invariant fields $A$ and $C$. Finally, the gauge-invariant field $E$ is obtained from Eq.~(\ref{Eq:Pert3bbis}).

Let us repeat that the above approach requires the knowledge of a static solution $({\cal F}_0(x),{\cal G}_0(x))$ of Eqs.~(\ref{Eq:WaveSystem},\ref{Eq:WaveConstraint}). A general strategy to obtain such a static solution is to make an infinitesimal variation along a static solution $(\alpha,\gamma,r,\Phi)$ of the Einstein-scalar equations with respect to its parameters. Since this linearization of the static solution automatically satisfies the linearized system~(\ref{Eq:Ev1}--\ref{Eq:Evm}), the corresponding gauge-invariant fields $A$ and $C$ (defined by Eqs.~(\ref{Eq:AGI},\ref{Eq:CGI})) fulfill the system~(\ref{Eq:Pert4bis},\ref{Eq:Pert5bis}), provided that we have the vanishing condition~(\ref{Eq:Pert3abis}) for $\sigma$; obviously, under the same condition, the fields $\cal F$ and $\cal G$ associated with $A$ and $C$ represent a static solution of the wave system~(\ref{Eq:WaveSystem},\ref{Eq:WaveConstraint}). In the following we will apply this general strategy to the cases of the Ellis-Bronnikov and AdS wormholes.

\subsection{Perturbed Ellis-Bronnikov wormhole}
\label{SubSec:PertEllis}

The Ellis-Bronnikov solution given in Eq.~(\ref{Eq:StaticSolutions})
can be linearized with respect to the parameters $b$ and $\gamma_1$, which yields
\begin{eqnarray}
\frac{\delta\gamma}{\gamma} &=&
- \left(\arctan\frac{x}{b} \right)\delta\gamma_1
 + \frac{\gamma_1 x}{x^2 + b^2}\delta b\,,\\
\frac{\delta r}{r} &=& -\left(\arctan  \frac{x}{b} \right)\delta\gamma_1
+ \frac{b + \gamma_1 x}{x^2 + b^2}\delta b\,,\\
\frac{\delta\Phi}{\Phi'} &=& \frac{\gamma_1}{1 + \gamma_1^2}(x^2 + b^2)
\left(\arctan  \frac{x}{b} \right)\frac{\delta\gamma_1}{b} - x\frac{\delta b}{b}\,.
\end{eqnarray}
Introduced into Eqs.~(\ref{Eq:AGI},\ref{Eq:CGI}) this gives rise to the gauge-invariant quantities
\begin{eqnarray}
A &=& -\frac{1}{1 + \gamma_1^2}\left[ \gamma_1
+ \left( 1 + 2\gamma_1\frac{x}{b} \right) \arctan \frac{x}{b} \right]
\delta\gamma_1 + \frac{\delta b}{b}\,,
\label{Eq:AEllis}\\
C &=& -\frac{1 + \gamma_1\frac{x}{b}}{1 + \gamma_1^2}\left(\arctan \frac{x}{b} \right) \delta\gamma_1
+ \frac{\delta b}{b}\, .
\label{Eq:CEllis}
\end{eqnarray}
As explained before, the fields $A$ and $C$ automatically satisfy the system of equations~(\ref{Eq:Pert3abis},\ref{Eq:Pert4bis},\ref{Eq:Pert5bis}). In this case the definition of $\sigma$ in Eq.~(\ref{Eq:Pert3abis}) gives $\sigma=-\delta b\gamma_1-b \delta \gamma_1=\delta(b\gamma_1)$, so that the condition $\sigma=0$ therein holds if
\begin{equation}
\delta b=-{b \delta\gamma_1\over \gamma_1}\,.
\label{Eq:SigmaEllis}
\end{equation}
Inserting Eqs.~(\ref{Eq:AEllis},\ref{Eq:CEllis},\ref{Eq:SigmaEllis}) into the definition~(\ref{Eq:FandG}) of $\cal F$ and $\cal G$ (and omitting the proportionality  factor $\delta\gamma_1$), one obtains the following time-independent solution of the constrained
wave system~(\ref{Eq:WaveSystem},\ref{Eq:WaveConstraint}):
\begin{equation}
\left( \begin{array}{cc} {\cal F}_0 \\ {\cal G}_0 \end{array} \right)
:= \frac{1}{r}\left( \begin{array}{c}
\frac{\gamma_1^2}{1 + \gamma_1^2}\left[ 1 + \frac{x}{b} \arctan\frac{x}{b} \right] \\
F(x)
\end{array} \right)\,,\qquad
F(x) := 1 + \frac{\gamma_1}{1 + \gamma_1^2}\left( 1 + \gamma_1\frac{x}{b} \right)
\arctan  \frac{x}{b}\,.
\label{Eq:StaticSolutionWaveSystem}
\end{equation}
Note that the function $F: \Real\to \Real$ is smooth and strictly positive. (\footnote{ For $\gamma_1 = 0$, $F = 1$ and the statement is trivial. When $\gamma_1\neq 0$ one has $F(x)\to +\infty$ for $x\to \pm\infty$, thus $F$ has a global minimum at some $x = x_0$, where $ 0 = (1 + \gamma_1^2) b F'(x_0) = \gamma_1\left[ \gamma_1\arctan(x_0/b) + (b + \gamma_1 x_0) b/(x_0^2 + b^2) \right]$. Eliminating the $\arctan$ term one obtains from this $(1 + \gamma_1^2) F(x_0) = (x_0 - b\gamma_1)^2/(x_0^2 + b^2)$. However, this minimum value must be strictly positive since otherwise $x_0 = b\gamma_1$ which would imply that $(1 + \gamma_1^2) b F'(x_0) = \gamma_1\left(\gamma_1\arctan\gamma_1 + 1 \right)$ which cannot be zero since $\gamma_1\neq 0$.})

Based on these observations, we can apply the general method for decoupling the
wave system~(\ref{Eq:WaveSystem},\ref{Eq:WaveConstraint}), choosing the static solution
$({\cal{F}}_0(x),{\cal{G}}_0(x))$ as in Eq.~(\ref{Eq:StaticSolutionWaveSystem}).
Note that in this case we have $Y_0 = 0$ and can choose $a = 1$ as $V = 0$,
which implies that $\mathcal{ V}=\tilde{\mathcal{ V}}$ and $\chi = \tilde{\chi}$.
One can verify that the function $\frac{\alpha^2}{\gamma^2} \mathcal{ V}$ appearing
in the master equation~(\ref{Eq:Master2}) agrees up to a rescaling with the potential
defined in Eq.~(32) of Ref.~\cite{jGfGoS09a}, denoted therein with $W$; more precisely,
\begin{equation}
\label{Eq:W}
\left(\frac{\alpha^2}{\gamma^2} \mathcal{V}\right)(x) =~{1 \over b^2} W\!\left(\frac{x}{b}\right)\,.
\end{equation}

For future mention let us point out some features of this function,
following from the analysis of $W$ in \cite{jGfGoS09a}.
First of all $\frac{\alpha^2}{\gamma^2} \mathcal{V} : \Real \to \Real$ is a $C^\infty$ bounded function; moreover, if $\gamma_1 \neq 0$ one has
$\left(\frac{\alpha^2}{\gamma^2} \mathcal{ V}\right)(x) \sim 2 e^{\pm 2 \pi \gamma_1}/x^2$
for $x \mapsto \pm \infty$. In the reflection-symmetric case $\gamma_1=0$, where $\alpha=\gamma=1$, one obtains
\begin{equation}
\mathcal{V}(x) = - \frac{3 b^2}{(x^2 + b^2)^2} = - \frac{3 b^2}{r^4(x)}\,,
\label{Eq:Vsim}
\end{equation}
and the master equation~(\ref{Eq:Master2}) is found to coincide with
Eq.~(\ref{Eq:MasterEllis}).

Let us now sketch some spectral features of the Schr\"odinger operator $-\left({\alpha \over\gamma}{d\over dx}\right)^2  + {\alpha^2\over\gamma^2}\, \mathcal{V}$ appearing in the master equation~(\ref{Eq:Master2}) (which can be regarded as a selfadjoint operator in $L^2(\mathbb{R},{\gamma \over\alpha} dx)$); these features allow one to infer the linear instability of the Ellis-Bronnikov wormhole both for $\gamma_1\neq 0$ and for $\gamma_1=0$. As shown in~\cite{jGfGoS09a}, the \virg zero energy'' equation  $\left[-\left({\alpha\over\gamma}{d\over dx}\right)^2  + {\alpha^2\over\gamma^2}\, \mathcal{V} \right]\chi_0 = 0$ has a solution
\begin{equation}
\chi_0(x) = \frac{x - b\gamma_1}{r(x) F(x)}\,,
\label{Eq:EBZeroMode}
\end{equation}
which has precisely one zero in the interval $(-\infty,+\infty)$. According to the Sturm oscillation theorem (see for instance~\cite{Weid-Book}, \cite{bS05} and references therein) it follows that for each $\gamma_1$ including $\gamma_1 = 0$, the Schr\"odinger operator in the master equation possesses a single bound state with negative energy. Note that for $\gamma_1 \neq 0$ the function $\chi_0$ decays as $1/|x|$ for large $|x|$, so that it describes a bound state with zero energy, while for $\gamma_1 = 0$ it reduces to $\chi_0(x) = x/\sqrt{1 + x^2}$ which is not normalizable but still has a single zero. The existence of a single bound state with negative energy implies that the master equation~(\ref{Eq:Master2}) for each Ellis-Bronnikov wormhole possesses a unique mode diverging exponentially in time, a fact of course sufficient to infer the instability of the wormhole. In the next section we will give more details on the spectral properties of the Schr\"odinger operator and on the solution of the master equation~(\ref{Eq:Master2}) within a rigorous functional setting, also allowing comparison with the corresponding problem for the perturbed AdS wormhole.

\subsection{Perturbed AdS wormhole}
\label{SubSec:Pertads}

Next, we analyze the AdS wormhole in the coordinate system $(s,u)$, as described by Eq.~(\ref{Eq:StaticSolutionsPot2}) for arbitrary parameters $k,B > 0$, and apply the general framework presented in this section with $(s,u)$ in place of $(t,x)$. Although the static solution formally depends on two parameters $B$ and $k$, it is important to note that $k$ also appears in the potential function $V(\Phi)$ (see Eq.~(\ref{Eq:StaticSolutionsPot1})). However, since we regard the potential to be fixed in our perturbation analysis, we will exclude the possibility of varying $k$. In contrast to $k$, the parameter $B$ is free, and variation of the solution~(\ref{Eq:StaticSolutionsPot2}) with respect to it gives
\begin{eqnarray}
\label{Eq:Variations1}
\delta \alpha &=&\delta \gamma=0\,,\\
\label{Eq:Variations2}
{\frac{\delta r}{r}}&=&
\frac{2B\delta B}{1+2B^2-\cos u}\,,\\
\label{Eq:Variations3}
{\frac{\delta \Phi}{\Phi'}}&=&-\sin u\frac{ \delta B}{B(1+B^2)}\,.
\end{eqnarray}
Equations~(\ref{Eq:Variations1},\ref{Eq:Variations2},\ref{Eq:Variations3}), introduced into Eqs.~(\ref{Eq:AGI},\ref{Eq:CGI}), yields the following expressions for the gauge-invariant quantities $A$ and $C$:
\begin{equation}
A = \frac{1+\cos u}{2 B(1+ B^2)} \delta B\,,\qquad
C = \frac{{\delta B}}{B}\,.
\label{Eq:ABAdS}
\end{equation}
From here and from the definition of $\sigma$ in Eq.~(\ref{Eq:Pert3abis}) we see that $\sigma=0$, as required, for every choice of the perturbation $\delta B$. Inserting Eq.~(\ref{Eq:ABAdS}) into the definition~(\ref{Eq:FandG}) of $\cal F$ and $\cal G$ (and omitting the proportionality  factor $\delta B$) one obtains, also in this case, a static solution of the system~(\ref{Eq:WaveSystem},\ref{Eq:WaveConstraint}):
\begin{equation}
\left( \begin{array}{cc} {\cal F}_0 \\ {\cal G}_0 \end{array} \right)
:= \frac{\sqrt{2}k}{B} \cos\left( \frac{u}{2} \right)\times
\left( \begin{array}{c}
-\frac{\sqrt{1+2B^2-{\cos u}}}{2(1+B^2)} \\
\frac{1}{\sqrt{1+2B^2-{\cos u}}}
\end{array} \right)\,.
\label{Eq:SolStat2}
\end{equation}
Note that ${\cal G}_0$ is a strictly positive function of $u\in(-\pi,\pi)$, and that ${\cal F}_0/{\cal G}_0 = -(1 + 2B^2 - \cos u)/(2(1 + B^2))$.

Having found a nontrivial solution, we can now obtain the master equation governing the spherical symmetric linearized perturbations of the AdS wormhole, following the general method explained before. We observe that $Y_0 = -2\tan\frac{u}{2}$ and that we can choose the constants $a_0$ and $x_0$ in Eq.~(\ref{Eq:ChiTrans}) such that $a = 1/\alpha^2$; therefore Eq.~(\ref{Eq:ChiTrans}) reads
$$
\chi := \left({\cal F} - \frac{{\cal F}_0}{{\cal G}_0}{\cal G}\right)\alpha^2
$$
and the master equation~(\ref{Eq:Master2}) becomes (recalling that $\alpha/\gamma=1$ for the AdS wormhole in coordinates $(s,u)$)
\begin{equation}
\left[ \frac{\partial^2}{\partial s^2}
  - \frac{\partial^2}{\partial u^2}  +
 {\cal {V}} \right] \chi = 0\,,
 \label{Eq:MasterAdS}
\end{equation}
with the potential
\begin{equation}
{\mathcal{V}}(u) \equiv {\mathcal{V}}_B(u) = -\frac{B^2 \left(2+B^2+\cos u\right) }{\left(1+2 B^2-\cos u\right)^2}\,.
\label{Eq:PotentialB}
\end{equation}

For the following, we assume Dirichlet boundary conditions at the two asymptotic AdS ends, that is,
\begin{equation}
\chi(s,\pm \pi) = 0\,.
\label{Eq:BounCond}
\end{equation}
Since
$$
\chi = \frac{1}{\sqrt{2}\sqrt{1+2 B^2-\cos u}}\delta\gamma
- \frac{1 + \cos u}{4(1 + B^2)(1 + 2B^2-\cos u)} \delta r
$$
for $\delta \Phi=0$, a sufficient condition for Eq.~(\ref{Eq:BounCond}) to hold is that,
in the gauge $\delta \Phi=0$, the perturbed functions $\delta r$ and $\delta \gamma$
vanish at the far ends $u=\pm\pi$ of the wormhole. For general considerations on boundary conditions for field theories on AdS spaces, see~\cite{sAcIdS78}\cite{cK13}\cite{cDhFaM18}.

Now (following the same scheme of the previous subsection) let us sketch some spectral features of the Schr\"odinger operator $-d^2/u^2+\mathcal{V}(u)$ with Dirichlet boundary conditions at $u=\pm \pi$ (to be regarded as a selfadjoint operator in $L^2((-\pi,\pi),du)$); these facts will allow us to infer the linear instability of the AdS wormhole. The zero-energy Schr\"odinger equation $[-d^2/du^2 + \mathcal{V} ]\chi_0 = 0$ admits for each fixed $B > 0$ the general solution
\begin{equation}
\chi_0(u) = C_1 {\sin {u \over 2} \over \sqrt{1+2B^2 -\cos u}}
 + C_2 {-2 u \sin{u \over 2} + 4 B^2 \cos{u \over 2}\over \sqrt{1+2B^2 -\cos u}}\,,\qquad
-\pi < u < \pi\,,
\label{Eq:AdSZeroMode}
\end{equation}
with constants $C_1,C_2$. The Dirichlet boundary conditions $\chi_0(\pm\pi)=0$ are satisfied only in the trivial case $C_1=C_2=0$, which shows that none of these solutions is an eigenfunction of our Schr\"odinger operator. For $C_1 = -2\pi C_2\neq 0$ the zero-energy solution satisfies the left boundary condition, i.e. $\chi_0(-\pi) = 0$, and since this solution has precisely one zero in the interval $(-\pi,\pi)$, (\footnote{Let us justify this statement on the number of zeroes
of $\chi_0$ for the special choice $C_1 = -2\pi C_2\neq 0$. In this case we can write $\chi_0(u) =\left(-2 C_2  \cos \frac{u}{2}\right) w(u)/\sqrt{1+2 B^2-\cos u}$ where
$w : (-\pi,\pi)\to \Real$, $u \mapsto w(u) := { (u+\pi ) \tan \frac{u}{2}-2 B^2}$.
The zeroes of $\chi_0$ in $(-\pi,\pi)$ coincide with the zeroes of the function
$w$. To find the zeroes of $w$, it is useful to note that this function has derivative $w'(u) = \left({1\over 2} \sec ^2\frac{u}{2}\right)\left(u+\sin u+\pi\right)>0$ for all $u \in (-\pi,\pi)$; from $w'>0$ it follows that $w$ is a strictly monotonic bijection of $(-\pi,\pi)$ to $(-2 B^2-2, + \infty)$, and thus possesses a unique zero.
})
 it follows from the Sturm oscillation theorem (see Theorem~3.4 in~\cite{bS05}) that our Schr\"odinger operator (with Dirichlet boundary conditions) has a single bound state with negative energy $E < 0$. This state gives rise to an exponentially growing (in time) mode solution of the master equation~(\ref{Eq:MasterAdS}) which is proportional to $e^{\sqrt{-E} t}$; this establishes the linear instability of the AdS wormhole. In the next section we analyze the spectral properties of the Schr\"odinger operator and their implications for the solutions of the master equation in a rigorous functional setting.

\section{Spectral representation of the solutions of the master equations}
\label{Sec:Spectral}

In this section we provide a rigorous analysis regarding the spectral properties of the Schr\"odinger-type operators involved in the master equations discussed so far. The next subsection and the related Appendix~\ref{appebro} concern the reflection-symmetric Ellis-Bronnikov case; the subsequent subsection sketches a similar analysis for the nonsymmetric case. The third subsection and the related Appendix~\ref{appeads} treat the corresponding problem for the AdS case (not previously considered in the literature, to the best of our knowledge); in the same subsection we establish bounds on the eigenvalues of the Schr\"odinger operator. Brief comments regarding the timescale associated with the instability are made in the final subsection.

\subsection{Spectral decomposition of the master equation and instability of the Ellis-Bronnikov wormhole in the reflection-symmetric case}
\label{Sec:SpectralA}
Let us consider the reflection-symmetric Ellis-Bronnikov wormhole and the corresponding master equation~(\ref{Eq:MasterEllis}), containing the potential $\mathcal{V}(x) = - 3 b^2/(x^2 + b^2)^2$; this equation can be written as
\begin{equation}
\ddot{\chi}(t) + \Hop \chi(t) = 0 \quad (t\in \Real),
\label{masterbro}
\end{equation}
where $\chi(t)$ stands for the function $\Real \ni x \mapsto \chi(t,x)$, and $\Hop$ indicates the operator $- d^2/d x^2 + \mathcal{V}$.
If we want a rigorous functional setting for Eq.~(\ref{masterbro}), we are led to consider
the Hilbert space
(\footnote{Throughout the paper, the expression ``Hilbert space'' is an abbreviation
for ``complex, separable Hilbert space''.})
\begin{equation}
\label{hilbro}
\Hilb := L^2(\Real, d x)
\end{equation}
made of the functions $f: \Real \to \Complex$, $x \mapsto f(x)$ which
are square integrable for the Lebesgue measure $d x$; we will write $\langle ~|~\rangle$
and $\|~\|$ for the natural inner product and norm of this space, defined by
$\langle f | \ell \rangle := \int_{\Real} d x \bar{f}(x) \ell(x)$
and $\| f \|^2 = \langle f | f \rangle$ for $f, \ell \in \Hilb$.
$\Hop$ can be regarded
as a selfadjoint operator in $\Hilb$, if we give for it the precise definition
\begin{equation}
\label{hbro}
\Hop := -\frac{d^2}{d x^2} + \mathcal{V} : \mathfrak{D} \subset \Hilb
\to \Hilb\,, \quad
\mathfrak{D} := \{ f \in \Hilb~|~f_{x x} \in \Hilb \}
\end{equation}
intending all $x$-derivatives in the distributional sense
({\footnote{The conditions $f \in \Hilb$ and $f_{x x} \in \Hilb$
imply $f_x \in \Hilb$, due to the Gagliardo-Nirenberg interpolation inequality
(see e.g. \cite{Adams-Book});
$\mathfrak{D}$ is just the usual Sobolev space $W^{2, 2}(\Real) \equiv H^2(\Real)$.
Let us also remark that, for $f \in \Hilb$, one has automatically $\mathcal{V} f \in \Hilb$
due to the boundedness of $\mathcal{V}$.}}).
Due to general facts on Schr\"odinger operators~\cite{Ber}, and to a specific analysis performed in~\cite{jGfGoS09a} for the potential $\mathcal{V}(x) = - 3 b^2/(x^2 + b^2)^2$, we can state
that the spectrum of $\Hop$ is the union of:
\begin{itemize}
\setlength\itemsep{-0.1cm}
\item[(i)] the point spectrum, which consists of a unique, simple eigenvalue $\mu_1 < 0$; \\
\item[(ii)] the continuous spectrum $[0,+\infty)$.
\end{itemize}
One can construct a generalized orthonormal basis
of the Hilbert space $\Hilb$, in the sense explained by Appendix \ref{appebro} and by \cite{Ber}, using:
\begin{itemize}
\setlength\itemsep{-0.1cm}
\item[(i)] a normalized eigenfunction $e_1$ for the eigenvalue
$\mu_1$ ($e_1 \in \mathfrak{D}$, $\Hop e_1 = \mu_1 e_1$, $\| e_1 \|=1$;
$e_1$ is proved to be $C^\infty$);
\item[(ii)] two suitably chosen ``improper eigenfunctions'' $e_{i \lambda}$ ($i=1,2$) for each
$\lambda \in (0,+\infty)$ (i.e., for each nonzero point $\lambda$ of the continuous spectrum); these are two linearly independent $C^\infty$ functions on $\Real$ which fulfill
$- d^2 e_{i \lambda}/ d x^2 + \mathcal{V} e_{i \lambda} = \lambda e_{i \lambda}$ but do not belong to
$\Hilb$.
\end{itemize}
Then, one can search for the solution
$\Real \ni t \to \chi(t)$ of
Eq.~(\ref{masterbro}) with appropriate smoothness properties and with
the initial conditions
\begin{equation} \chi(0) = q\,, \quad  \dot{\chi}(0) = p\,,
\label{cond}
\end{equation}
where $q: x \mapsto q(x)$ and $p : x \mapsto p(x)$ are sufficiently regular functions.
For all technical details, we
refer again to Appendix \ref{appebro}; here we introduce
the selfadjoint operator $| \Hop |^{1/2}: \mathfrak{D}^{1/2}
\subset \Hilb \to \Hilb$ and indicate how to regard
the domains $\mathfrak{D}^{1/2}$ and $\mathfrak{D}$ as Hilbert spaces with their
own inner products. One can show that, for any $q \in \mathfrak{D}$ and $p \in \mathfrak{D}^{1/2}$, Eqs.~(\ref{masterbro},\ref{cond}) have a unique solution $t \mapsto \chi(t)$ in
$C(\Real, \mathfrak{D}) \cap C^1(\Real, \mathfrak{D}^{1/2}) \cap C^2(\Real, \Hilb)$,
which is as follows for all $t \in \Real$:
\begin{equation}
\chi(t) = \left[ \langle e_1 | q \rangle \cosh( |\mu_1|^{1/2} t) +
\langle e_1 | p \rangle \frac{\sinh( |\mu_1|^{1/2} t)}{|\mu_1|^{1/2}}  \right] e_1 +
\sum_{i=1}^{2} \int_{0}^{+\infty} \hspace{-0.4cm} d \lambda
\left[ \langle e_{i \lambda} | q \rangle \cos( \lambda^{1/2} t) +
\langle e_{i \lambda} | p \rangle \frac{\sin( \lambda^{1/2} t)}{\lambda^{1/2}}  \right] e_{i \lambda}\, .
\label{solbro}
\end{equation}
As explained in Appendix~\ref{appebro}, the symbols $\langle \cdot | \cdot \rangle$ in the above formula indicate usual inner products in $\Hilb$, or suitably defined generalizations;
the integrals over $\lambda$ are understood in a weak sense. Of course, we are interested
in the case where $\chi(t)$ is \textsl{real valued} for each $t$, which occurs if
and only if the data $q, p$ are real-valued functions.

The coefficient of $e_1$ in Eq.~(\ref{solbro}) diverges exponentially both for
$t \to -\infty$ and for $t \to + \infty$ (except for very special choices of
$\langle e_1 | q \rangle$ and $\langle e_1 | p \rangle$
 (\footnote{\label{footnote13}For $\langle e_1 | q \rangle = \langle e_1 | p \rangle = 0$,
the coefficient of $e_1$ in~(\ref{solbro}) vanishes. For
$\langle e_1 | q \rangle = \xi \langle e_1 | p \rangle/|\mu_1|^{1/2} \neq 0$,
with $\xi = \pm 1$, the coefficient of $e_1$ diverges for $t \to \xi (+\infty)$
and vanishes for $t \to \xi (-\infty)$.})); this suffices to infer the (linear) instability
of the reflection-symmetric Ellis-Bronnikov wormhole
\cite{jGfGoS08} \cite{jGfGoS09a} \cite{fCfPlP19}. In addition, let us remark that the integrals
over $\lambda$ in Eq.~(\ref{solbro}) are superpositions of ``non-normalizable'' oscillatory modes, living outside the space $\Hilb = L^2(\Real, d x)$ like the improper eigenfunctions $e_{i \lambda}$.

\subsection{Spectral decomposition of the master equation and instability of the Ellis-Bronnikov wormhole in the nonsymmetric case}
\label{Sec:SpectralB}

Let us now pass to the non-reflection-symmetric Ellis-Bronnikov wormhole
(as in Eq.~(\ref{Eq:StaticSolutions}) with $\gamma_1 \neq 0$). In this case the master equation for $\chi(t,x)$ has the form~(\ref{Eq:Master2}), involving the operator
\begin{equation}
- \left(\frac{\alpha}{\gamma} \frac{\partial}{\partial x}\right)^2  + \frac{\alpha^2}{\gamma^2} \mathcal{V}\,, \quad
\frac{\alpha(x)}{\gamma(x)} = e^{2 \gamma_1 \arctan \frac{x}{b}}~~(x \in \Real)\,.
\end{equation}
As noted in \cite{jGfGoS09a}, the spectral analysis of this case can be simplified by
introducing the new coordinate
\begin{equation}
\label{Eq:ro}
\rho = \rho(x) := \int_{0}^x \frac{\gamma(y)}{\alpha(y)} \,dy\,;
\end{equation}
note that the mapping $x \mapsto \rho(x)$ is a diffeomorphism of $\Real$
to itself, and $\rho(x) \sim e^{\mp \pi \gamma_1} x$ for $x \to \pm \infty$.
By construction
$\frac{\alpha}{\gamma} \frac{\partial}{\partial x}$ $= \frac{\partial}{\partial \rho}$;
so, by writing $\chi(t,\rho)$ as an abbreviation for $\chi(t,x(\rho))$ we can
rephrase the master equation~(\ref{Eq:Master2}) as
\begin{equation}
\label{Eq:Master2reph}
\left[ \frac{\partial^2}{\partial t^2}
   - \left( \frac{\partial}{\partial \rho} \right)^2
 + \mathcal{U}(\rho) \right] \chi(t, \rho) = 0\,,
\quad \mathcal{U}(\rho) := \left(\frac{\alpha^2}{\gamma^2} \mathcal{V}\right)(x(\rho))
\quad (t,\rho \in \Real).
\end{equation}
The function $\mathcal{U} : \Real \to \Real$ is $C^\infty$;
due to the $x \to \pm \infty$ asymptotics of
$\rho(x)$ (see after Eq.~(\ref{Eq:ro})) and $\left(\frac{\alpha^2}{\gamma^2} \mathcal{V}\right)(x)$
(see after Eq.~(\ref{Eq:W})), we have $\mathcal{U}(\rho) \sim 2/\rho^2$ for $\rho \to \pm \infty$.
A precise functional setting for Eq.~(\ref{Eq:Master2reph}) can be obtained by introducing the
Hilbert space and the selfadjoint operator
\begin{equation}
\label{hbrons}
\Hilb := L^2(\Real, d \rho)\,; \qquad
\Hop := -\frac{d^2}{d \rho^2} + \mathcal{U} : \mathfrak{D} \subset \Hilb
\to \Hilb\,, \quad
\mathfrak{D} := \{ f \in \Hilb~|~f_{\rho \rho} \in \Hilb \}
\end{equation}
(the $\rho$-derivatives are meant distributionally);
$\langle~|~\rangle$ and $\|~\|$ indicate in the sequel
the natural inner product and norm of $\Hilb$. (\footnote{\label{footnote:Hilbert} Note that, since $d\rho={\gamma\over \alpha}dx$, working with the operator and the Hilbert space defined in Eq.~(\ref{hbrons}) is equivalent to working directly with the operator $-\left({\alpha\over\gamma} {d\over d x}\right)^2+{\alpha^2\over\gamma^2}\mathcal{ V}$ in the Hilbert space $L^2(\mathbb{R},{\gamma\over\alpha} dx)$, which is the formulation considered at the end of subsection \ref{SubSec:PertEllis}.})
After giving these prescriptions we write Eq.~(\ref{Eq:Master2reph}) in the form~(\ref{masterbro}),
where $\chi(t)$ stands for the function $\rho \mapsto \chi(t,\rho)$; obviously enough,
the treatment of this equation is reduced to a spectral analysis of the
Schr\"odinger operator $\Hop$ in Eq.~(\ref{hbrons}), which is rather
similar to the discussion of the operator~(\ref{hbro}) for the reflection-symmetric wormhole.

The main difference with respect to the symmetric case  is that
the operator $\Hop$ in Eq.~(\ref{hbrons}) has a point spectrum
consisting of \textsl{two} simple eigenvalues $\mu_1 < 0$ and $\mu_2 := 0$, see the comments below Eq.~(\ref{Eq:EBZeroMode}); the continuous spectrum is $(0,+\infty)$. Due to these facts there is a generalized orthonormal basis
made of normalized eigenfunctions $e_1, e_2$ for the eigenvalues $\mu_1 < 0$
and $\mu_2 = 0$ ($e_1, e_2 \in \mathfrak{D}(\Hop)$, $\Hop e_1 = \mu_1 e_1$,
$\Hop e_2 = 0$, $\| e_1 \| = \| e_2 \| = 1$),
plus two improper eigenfunctions $e_{i \lambda}$ ($i=1,2$)
for each $\lambda$ in the continuous spectrum.

As in the symmetric case, one can define Hilbert space
structures for the domains $\mathfrak{D}$, $\mathfrak{D}^{1/2}$ of the operators
$\Hop$, $|\Hop|^{1/2}$. For $q \in \mathfrak{D}$ and $p \in \mathfrak{D}^{1/2}$,
the master equation ~(\ref{masterbro}) with initial conditions~(\ref{cond}) is proved
again to possess a unique solution $t \mapsto \chi(t)$ in
$C(\Real, \mathfrak{D}) \cap C^1(\Real, \mathfrak{D}^{1/2}) \cap C^2(\Real, \Hilb)$;
this has a representation similar to~(\ref{solbro}) with an additional
term associated with the eigenvalue zero, namely:
\begin{equation}
\chi(t) = \mbox{r.h.s. of Eq.~(\ref{solbro})} +
\Big[ \langle e_2 | q \rangle + \langle e_2 | p \rangle t  \Big] e_2\,.
\label{solbrons}
\end{equation}
So, besides the exponentially divergent term proportional
to $e_1$, the expression of $\chi(t)$ contains a term diverging
linearly for $t \to \pm \infty$ (if $\langle e_2 | p \rangle \neq 0$);
in any case the wormhole is
linearly unstable. Let us note that, as in Eq.~(\ref{solbro}),
the present expression for $\chi(t)$ contains an integral
over $\lambda$ of non-normalizable oscillatory modes,
proportional to the improper eigenfunctions $e_{i \lambda}$ which
live outside $\Hilb$.

\subsection{Spectral decomposition of the master equation and instability of the AdS wormhole}

In the AdS case we introduce the Hilbert space
\begin{equation}
\label{hilbu}
\Hilb := L^2((-\pi,\pi), du)
\end{equation}
formed by the functions $f : (-\pi,\pi) \to \Complex$, $u \mapsto f(u)$ which
are square integrable with respect to the Lebesgue measure $d u$;  from now on we
denote by $\Braket{~|~}$ and $\|~\|$ the natural inner product and norm of $\Hilb$,
so that $\langle f | \ell \rangle := \int_{-\pi,\pi} d u \bar{f}(u) \ell(u)$
and $\| f \|^2 = \langle f | f \rangle$ for $f, \ell \in \Hilb$.
In addition,
let us consider the potential
${\mathcal{ V} }$ appearing in Eq.~(\ref{Eq:PotentialB}) (a $C^\infty$ function on
$[-\pi, \pi]$). A rigorous setting for the master equation~(\ref{Eq:MasterAdS}) with boundary conditions~(\ref{Eq:BounCond}) can be set up using the space~(\ref{hilbu}) and the selfadjoint operator
\begin{equation}
\label{hB}
\Hop := -{d^2 \over du^2}+ \mathcal{ V} : \mathfrak{D} \subset \Hilb \to \Hilb\,,
\quad
\mathfrak{D} := \{ f \in \Hilb~|~f_{u u} \in \Hilb~, f(\pm \pi) = 0 \}\,.
\end{equation}
Here and in the sequel, the $u$-derivatives like $f_{u u }$ are understood distributionally;
a function $f\in \Hilb$ with $f_{uu}\in \Hilb$ is in fact in $C^1([-\pi,\pi])$,
so it can be evaluated at $u=\pm\pi$
(\footnote{The conditions $f \in \Hilb$, $f_{u u} \in \Hilb$
imply $f_u \in \Hilb$, due to the already mentioned Gagliardo-Nirenberg interpolation inequality
\cite{Adams-Book}.
The space $\{ f \in \Hilb~|~f_{u u} \in \Hilb \}$ coincides with
the standard Sobolev space $W^{2, 2}(-\pi,\pi) \equiv H^{2}(-\pi,\pi)$, which is contained in $C^1([-\pi,\pi])$
by the Sobolev embedding theorem (see again \cite{Adams-Book}). Let us also remark that,
due to the boundedness of the function $\mathcal{V}$, for each $f \in \Hilb$
one has automatically $\mathcal{V} f \in \Hilb$.}).
As an operator in the Hilbert space $\Hilb$, $\Hop$ has the following properties:
\begin{itemize}
\setlength\itemsep{-0.1cm}
\item[(i)] it is selfadjoint;
\item[(ii)] it is bounded from below;
\item[(iii)] it has a purely discrete spectrum.
\end{itemize}
\noindent
As known in general for Hilbert space operators
satisfying properties (i-iii), it is possible to represent the eigenvalues of $\Hop$ as an increasing sequence $\mu_1 <\mu_2 < \cdots$. In addition, $\Hop$ has the following properties:
\begin{itemize}
\setlength\itemsep{-0.1cm}
\item[(iv)] any of its eigenfunctions is in the space $C^{\infty}([-\pi,\pi])$;
\item[(v)] each one of its eigenvalues is simple.
\end{itemize}
\noindent
For future mention, let us recall that the operator $\Hop^0:=-d^2/du^2$ with domain $\mathfrak{D}$ as above also has
the properties (i-v); in this case the eigenvalues are $\mu^{0}_n :=n^2/4$, with normalized
eigenfunctions $f^0_n(u):=(1/\sqrt{\pi})\sin[(n/2)(u+\pi)]$ ($n=1,2,\ldots$)
 (\footnote{Let us give more complete information on the above issues (i-v).
For some general facts about Hilbert space operators with properties (i-iii)
(including the possibility to arrange their eigenvalues in an increasing sequence),
see e.g. \cite{Schmudgen-Book} (especially, pages 37, 178 and 265-67).
To go on, let us recall the following regularity result: if $f$ is a distribution on an open interval $\Omega \subset \Real$ (with derivatives $f^{(i)}$, $i=0,1,\ldots$) and $f$ fulfills a homogeneous linear ODE
$f^{(k)} + \sum_{i=0}^{k-1} a_i f^{(i)}=0$ of any order
$k \in \{1,2,\ldots\}$ with $C^\infty$ coefficients
$a_i: \Omega \to \Complex$, then $f$ is a $C^\infty$ function on $\Omega$:
this follows from Theorem IX in \cite{Schwartz-Book}, page 130.
The properties (i-v) of $\Hop^0$ and the expressions given above
for its eigenvalues and eigenfunctions are checked ``by hand'', keeping
in mind that the eigenfunctions are smooth
due to the previously mentioned regularity result.
Now consider any function $\mathcal{V}\in C^{\infty}([-\pi,\pi],\mathbb{R})$;
then, due to the boundedness of this function,
the multiplication operator by $\cal V$ is a bounded selfadjoint operator on $\Hilb$.
As well known the properties (i), or (i-ii), or (i-iii) of an operator in an abstract Hilbert space
are preserved by the addition
of a bounded selfadjoint perturbation (see again \cite{Schmudgen-Book}); therefore the operator
$\Hop := \Hop^0+\mathcal{V}=-d^2/du^2+\mathcal{V}$ with domain $\mathfrak{D}$
fulfills (i-iii). The operator $\Hop$ also has the
properties (iv-v). For the proof of (iv) one can use again the cited
regularity result for distributional, homogeneous linear ODEs;
a derivation of (v) can be found e.g. in \cite{Poschel-Book}, page 30.
All the previous statements apply, in particular, with ${\cal V}$ as in Eq.~(\ref{Eq:PotentialB}).
}).

In the remainder of this section the notations $\mathcal{V}$, $\Hilb$, $\mathfrak{D}$, $\Hop$, $(\mu_n)_{n=1,2,\ldots}$ will always indicate, respectively, the potential $\mathcal{V}$ in Eq.~(\ref{Eq:PotentialB}), the Hilbert space in Eq.~(\ref{hilbu}), the domain and the operator in Eq.~(\ref{hB}), and the eigenvalues of this operator in increasing order. Sometimes it will be useful to emphasize that the potential $\mathcal{V}$ depends on the parameter $B \in (0,+\infty)$, thus originating in a similar dependence for the corresponding operator and its eigenvalues: $\mathcal{V} \equiv \mathcal{V}_B$, $\Hop \equiv {\Hop}_B$, $\mu_n \equiv \mu_n(B)$ ($n=1,2,\ldots$).

As discussed in section~\ref{SubSec:Pertads}, the analysis of the zero-energy solutions in Eq.~(\ref{Eq:AdSZeroMode}) implies that the ground-state energy is negative, while all other eigenvalues are positive, such that
\begin{equation}
\label{signs}
\mu_1 < 0 < \mu_2 < \mu_3 < \cdots\,,
\end{equation}
the negative eigenvalue $\mu_1$ being associated a mode of the master equation~(\ref{Eq:MasterAdS}) growing exponentially in time, whereas in contrast to this, the eigenvalues $\mu_n$ for $n\geqslant 2$ are associated with oscillatory modes.

In what follows, we provide estimates for the eigenvalues of $\mu_n(B)$. We start with an upper bound for the ground-state energy $\mu_1 \equiv \mu_1(B)$. According to the Rayleigh-Ritz variational characterization
(see e.g. \cite{Schmudgen-Book}, pages 265-266) one has
\begin{equation}
\mu_1(B)=\inf_{f \in \mathfrak{D} \setminus \{0\}} {\Braket{f|\Hop_B f} \over||f||^2} \, .
\label{Eq:InfSpectrum}
\end{equation}
Choosing in $\mathfrak{D}$ the function
\begin{equation}
 f(u) := \cos{u \over 2},
 \end{equation}
we get
\begin{equation}
{\Braket{f|\Hop_B f} \over||f||^2}=\frac{1}{4}-B^2+\frac{\sqrt{1 + B^2} \left(4 B^2 -3\right)}{4B}
=: \varepsilon(B)\,,
\label{Eq:epsilonB}
\end{equation}
which, together with Eq.~(\ref{Eq:InfSpectrum}), yields the estimate
\begin{equation*}
    \mu_1(B)\leqslant \epsilon(B)
\end{equation*}
for each $B>0$. It can be checked that $B\mapsto\varepsilon(B)$ is a negative, monotonously increasing function on $(0,+\infty)$ with the properties
\begin{equation}
\lim\limits_{B\to 0^+} \varepsilon(B) = -\infty\,,\qquad
\lim\limits_{B\to +\infty} \varepsilon(B) = 0^-\,.
\end{equation}
Therefore, we obtain the upper bound for the ground-state energy
\begin{equation}
\mu_1(B)\leqslant \varepsilon(B) < 0
\label{Eq:mu1Bound}
\end{equation}
which provides an independent proof for the fact that it is negative, and hence also for the linear instability of the AdS wormhole.

Next, we provide two-sided bounds on the eigenvalues $\mu_n \equiv \mu_n(B)$ for arbitrary $n$. In order to achieve this, we check that for any fixed $B > 0$, one has
\begin{equation}
\min_{u \in [-\pi,\pi]} {\mathcal{V}}_B(u) = {\mathcal{V}}_B(0) = - {1 \over 4} - {3 \over 4 B^2}\,,
\qquad \max_{u \in [-\pi,\pi]} {\mathcal{V}}_B(u)
=  {\mathcal{V}}_B(\pm \pi) = - {1 \over 4} + {1 \over 4 (1 + B^2)}\,.
\label{Eq:LimPot}
 \end{equation}
In the Hilbert space $\Hilb$, let us consider the operators
$\Hop =-{d^2\over du^2}+{\cal V}$, $\Hop^- :=-{d^2\over du^2} - {1 \over 4} - {3 \over 4 B^2}$
and $\Hop^+ :=-{d^2\over du^2} - {1 \over 4} + {1 \over 4 (1 + B^2)}$,
all of them with the same domain $\mathfrak{D}$ as defined in Eq.~(\ref{hB})
(and all of them satisfying the properties (i-v) after the cited equation). Due to Eq.~(\ref{Eq:LimPot}) we have
$\Braket{f|\Hop^- f}\leqslant \Braket{f|\Hop f}\leqslant\Braket{f|\Hop^+ f}$
for all $f\in \mathfrak{D}$, and this implies (see e.g. \cite{Schmudgen-Book}, pages 230 and 267)
$\mu^{-}_n \leqslant \mu_n \leqslant \mu^{+}_n$ for $n=1,2,\ldots$,
where $\mu^{\mp}_1 < \mu^{\mp}_2 < \cdots$ are the eigenvalues of $\Hop^{\mp}$.
On the other hand, the eigenvalues of $\Hop^{\mp}$ are obtained by shifting those
of $\Hop^0=-d^2/du^2$, i.e., $\mu^{-}_n ={n^2 \over 4} - {1 \over 4} - {3 \over 4 B^2}$
and $\mu^{+}_n ={n^2 \over 4} - {1 \over 4} + {1 \over 4 (1 + B^2)}$. In conclusion, the eigenvalues of $\Hop$ satisfy the two-sided bounds
\begin{equation}
\label{Eq:boundsmun}
{n^2 - 1 \over 4}  - {3 \over 4 B^2} \leqslant \mu_n(B)  \leqslant {n^2 - 1\over 4}
+ {1 \over 4 (1 + B^2)} \qquad (n=1,2,\ldots)\,.
\end{equation}
Combining this result with Eq.~(\ref{Eq:mu1Bound}) one obtains the following two-sided bound for the ground-state energy:
\begin{equation}
\label{Eq:boundsmu1}
-\frac{3}{4B^2} \leqslant \mu_1(B) \leqslant \varepsilon(B) = -\frac{1}{2B^2} + {\cal O}\left( \frac{1}{B^4} \right)\,.
\end{equation}

After these remarks concerning the eigenvalues of the Schr\"odinger operator $\Hop$, we discuss the spectral decomposition of the master equation. To this purpose, we choose
for each $n$ a normalized eigenfunction $e_n$ for the (simple) eigenvalue $\mu_n$:
\begin{equation}
\label{basisads}
e_n \in \mathfrak{D}~, \quad \Hop e_n = \mu_n e_n~, \quad \| e_n \| = 1 \qquad (n=1,2,\ldots) \, .
\end{equation}
Then $(e_n)_{n=1,2,\ldots}$ is an orthonormal basis of $\Hilb$ (in the ordinary sense),
due to the spectral theorem for selfadjoint operators with a purely discrete spectrum. In comparison with the previous analysis for the Ellis-Bronnikov wormhole, we do not have the technical complications associated with the continuous spectrum and to the related ``improper'' eigenfunctions.

Next, we write the master equation~(\ref{Eq:MasterAdS}) in a form similar to~(\ref{masterbro}) and add initial conditions as in~(\ref{cond}); in this way we obtain the system
\begin{equation}
\ddot{\chi}(s) + \Hop \chi(s) = 0 ~~  (s \in \Real)\,,
\quad \chi(0)=q\,,\quad \dot{\chi}(0) = p\,,
\label{masterads}
\end{equation}
where $\chi(s)$ refers to the function $u \mapsto \chi(s,u)$, the dots stand
for $s$-derivatives and $q: u \mapsto q(u)$, $p: u \mapsto p(u)$ are functions with
appropriate regularity.

A technically precise framework for the discussion of the system~(\ref{masterads})
is provided by Appendix \ref{appeads} where we introduce
(similarly to the previous treatment for the Ellis-Bronnikov wormhole)
the selfadjoint operator $| \Hop |^{1/2}: \mathfrak{D}^{1/2}
\subset \Hilb \to \Hilb$ and indicate how to regard
the domains $\mathfrak{D}^{1/2}$ and $\mathfrak{D}$ as Hilbert spaces with appropriate
inner products. It turns out that, for any $q \in \mathfrak{D}$ and $p \in \mathfrak{D}^{1/2}$,
the system~(\ref{masterads}) has a unique solution $s \mapsto \chi(s)$ in
$C(\Real, \mathfrak{D}) \cap C^1(\Real, \mathfrak{D}^{1/2}) \cap C^2(\Real, \Hilb)$;
using an orthonormal basis $(e_n)_{n=1,2,\ldots}$ as in Eq.~(\ref{basisads}), the solution
can be written as follows
for all $s \in \Real$:
\begin{equation}
\chi(s) = \left[ \langle e_1 | q \rangle \cosh( |\mu_1|^{1/2} s) +
\langle e_1 | p \rangle \frac{\sinh( |\mu_1|^{1/2} s)}{|\mu_1|^{1/2}}  \right] e_1 +
\sum_{n=2}^{+\infty}
\left[ \langle e_{n} | q \rangle \cos( \mu^{1/2}_n s) +
\langle e_{n} | p \rangle \frac{\sin( \mu^{1/2}_n s)}{\mu^{1/2}_n}  \right] e_{n}\, .
\label{solads}
\end{equation}
The above function $\chi(s)$ is real valued for each $s$ if
and only if the data $q, p$ are real-valued functions.

The coefficient of $e_1$ in Eq.~(\ref{solads}) diverges exponentially both for
$s \to -\infty$, and for $s \to + \infty$ (except for very special choices of
$\langle e_1 | q \rangle$ and $\langle e_1 | p \rangle$
 (\footnote{See the footnote \ref{footnote13} in the discussion after
Eq.~(\ref{solbro}), which is readily adapted to the present framework.}));
so, the AdS wormhole is linearly unstable. For each $n \geqslant 2$, the $n$-th term
in Eq.~(\ref{solads}) represents a ``normalizable'' oscillatory mode, living
like $e_n$ inside the Hilbert space $\Hilb$ (indeed, inside the subspace $\mathfrak{D}\subset \Hilb$). This is a relevant difference with respect to the ``non-normalizable''
oscillatory modes that we have found for the perturbed Ellis-Bronnikov wormhole,
associated with the continuous spectrum and living outside the Hilbert space of the system
(see the comments after Eqs.~(\ref{solbro}) and~(\ref{solbrons})).

\subsection{Instability times}
\label{Sec:SpectralD}

In the Ellis-Bronnikov case it has been shown~\cite{jGfGoS09a} that the timescale $\tau_\text{unstable}$ (measured with respect to proper time at the throat of the unperturbed solution) associated with the unstable mode is of the order of the throat's areal radius $r_\text{throat}$ divided by the speed of light. The estimates provided in Eq.~(\ref{Eq:boundsmu1}) allow us to estimate the corresponding timescale for the AdS wormhole, and yield
\begin{equation}
\frac{1}{\sqrt{3}} \leq \frac{\tau_\text{unstable}}{r_\text{throat}}
 \leq \frac{1}{2B\sqrt{-\varepsilon(B)}}\,,
\end{equation}
with the function $\varepsilon(B)$ defined in Eq.~(\ref{Eq:epsilonB}). Since $2B\sqrt{-\varepsilon(B)} \to \sqrt{2}$ for large $B$ and since for $B\to 0$ the AdS wormhole reduces to the reflection-symmetric Ellis-Bronnikov wormhole (\footnote{See the comment on the limit $k\to 0$ after Eq. (\ref{Eq:StaticSolutionsPot1}), keeping in mind that $B=b k$.}), it follows also in this case that $\tau_\text{unstable}$ is of the order of the throat's areal radius (divided by the speed of light in physical units).

\section{A dS wormhole with horizons and its linearized perturbations}
\label{Section:dS}

Let us return to the Bronnikov-Fabris wormhole solution mentioned at the beginning of section \ref{Subsection:AdS}, depending on the parameters $M$ and $K$. Keeping the assumption (\ref{Eq:ConstantM}) that $M=0$ we can as well consider, as an alternative to (\ref{Eq:ConstantK}), the choice
\begin{equation}
K\equiv k^2\,, \qquad (k>0)\,.
\end{equation}
In this way we obtain
\begin{equation}
V(\Phi)= \frac{k^2}{\kappa}
\left[ 3 -2  \cos ^2\left( \sqrt{\frac{\kappa}{2}}\Phi\right) \right]\,,\quad
\alpha = \gamma^{-1} = \sqrt{1 - k^2(x^2+b^2)}\,,\quad
r = \sqrt{x^2 + b^2}\,,\quad
\Phi = \sqrt{\frac{2}{\kappa}}\arctan\frac{x}{b}\,.
\label{Eq:StaticSolutionsPotdS}
\end{equation}
For $b\to0$ the third equality in~(\ref{Eq:StaticSolutionsPotdS}) should be read as $r=x>0$, and the corresponding metric represents a dS universe with cosmological constant $\Lambda = 3k^2$. From now on we intend
\begin{equation}
b\in\left(0,{1\over k}\right)\,;
\label{b}
\end{equation}
the limitation $b<{1\over k}$ ensures that the expressions for $\alpha$ and $\gamma$ in Eq. (\ref{Eq:StaticSolutionsPotdS}), if taken literally, make sense near the throat $x=0$ or, more substantially, that $\partial_t$ is actually timelike and $\partial_x$ is actually spacelike near $x=0$.
We also set
\begin{equation}
B:=b\, k\in(0,1)\,,\qquad\ell:={\sqrt{1-B^2}\over k}\,;
\label{Eq:B_ell}
\end{equation}
the metric corresponding to the above coefficients reads
\begin{align}
{\bf g}& = - \left[ 1 - k^2(x^2+b^2) \right] dt^2
+ \frac{d x^2}{ 1 - k^2(x^2+b^2)}
+ \left( x^2+b^2\right) (d \vartheta^2 + \sin^2 \vartheta\, d \varphi^2)
\nonumber\\
&=
-(1-B^2) \left( 1 - \frac{x^2}{\ell^2} \right) dt^2
+ \frac{d x^2}{(1-B^2)(1 - \frac{x^2}{\ell^2})}
+ \left( x^2 + b^2 \right) (d \vartheta^2 + \sin^2 \vartheta\, d \varphi^2)\,.
\label{Eq:dSWH1}
\end{align}
By analogy with the terminology of section
\ref{Subsection:AdS}, we refer to this as a ``dS wormhole''; let us note that the
expressions for $\Phi$, $V(\Phi)$, $r$ in Eq.~\eqref{Eq:StaticSolutionsPotdS} and
the expression \eqref{Eq:dSWH1} for ${\bf g}$ can be obtained formally from
the analogous expressions of the AdS case (see Eq.~\eqref{Eq:StaticSolutionsPot1}) by making the replacement $k \mapsto i k$.

Let us consider the regions
\begin{equation*}
I:=\{(t,x)\,|\,t\in \mathbb{R}\,,\, x \in(-\ell, \ell)\}\,,
\end{equation*}
\begin{equation}
E^-:=\{(t,x)\,|\,t\in \mathbb{R}\,,\, x \in(-\infty, -\ell)\}\,,\qquad E^+:=\{(t,x)\,|\,t\in \mathbb{R}\,,\, x \in(\ell, +\infty) \}\,;
\label{region_I_E}
\end{equation}
then the expressions for $\alpha$ and $\gamma$ in Eq.(\ref{Eq:StaticSolutionsPotdS}) are well defined in a literal sense over $I$; more substantially, the metric \eqref{Eq:dSWH1} is well defined over $I \times S^2$ and the vector fields $\partial_t$ and $\partial_x$ are, respectively, timelike and spacelike on this domain. However, Eq. (\ref{Eq:dSWH1}) also gives a Lorentzian metric on each one of the regions $E^-$ and $E^+$; here $\partial_t$ is spacelike and $\partial_x$ is timelike, so the metric is nonstatic. In the sequel we often refer to $I$ as the internal region and to $E^\pm$ as the exterior regions in $(t,x)$ space. At $x=\pm \ell$ the metric seems to be ill defined but, as explained hereafter, these are just apparent singularities related to the coordinate system: the hypersurfaces $x=\pm\ell$ are indeed cosmological horizons and the metric is nonsingular across them.
Let us note that the $b\to 0$ limit of the previous statement (with $x>0$) corresponds
to well-known features of the dS universe, having a horizon at $x=\ell={1\over k}$. \par
In the next paragraph we consider an alternative coordinatization for the
internal region $I$ introducing the analogs of the AdS wormhole coordinates $(s,u)$ (see Eq. (\ref{coordsu})); in the subsequent paragraphs we consider alternative parametrizations yielding a Kruskal-type extension of the metric (\ref{Eq:dSWH1}) which is regular across $x=\pm\ell$. The extended universe constructed in this way can also be interpreted as a regular black hole with an expanding cosmology beyond the horizons, and is hence referred to as a ``black universe'' in \cite{Bro2018}.

\subsection{Another coordinate system for the internal region $I$}

Let us put
\begin{equation}
t= {\ell\over 2\, (1-B^2)} \,s\,,\qquad x=\ell \,\mbox{tanh} {u \over 2}  \,,\qquad
(s, u)\in\Real^2  \,;
\label{coordsu_dS}
\end{equation}
the map $(s,u)\mapsto(t,x)$ is one to one between $\Real^2$ and the inner region $I$ (with inverse
$
s={2(1-B^2)\over \ell}\,t$, $ u=2\,\mbox{arctanh} {x\over\ell}= \log ({\ell+x\over\ell - x})
$).
We can regard $(s,u)$ as an alternative coordinate system for $I$; this does not eliminate the apparent singularities at $x = \pm \ell$ but sends them to infinity since the limits $x\to\pm \ell$ correspond to the limits $u\to\pm\infty$. In the new coordinates the metric \eqref{Eq:dSWH1} becomes
\begin{equation}
{\bf g} = \frac{1}{4 k^2 \cosh^2\frac{u}{2}}
\left[ -ds^2 + du^2
+2\bigg(\cosh u -(1-2 B^2)\bigg)
(d \vartheta^2 + \sin^2 \vartheta\, d \varphi^2) \right],
\label{gsu}
\end{equation}
with radial null geodesics given by the straight lines $s = \pm u + const$.

To conclude this paragraph, let us remark that the transformation \eqref{coordsu_dS} and the expression \eqref{gsu} for the metric can be obtained from their AdS analogs
(see Eqs.~(\ref{coordsu},\ref{Eq:StaticSolutionsPot2})) by
making the formal replacements $k \mapsto i k, B \mapsto i B, s \mapsto i s, u \mapsto i u$.

\subsection{A first spacetime extension}

We start our construction from the internal region $I$, that we
describe in terms of the coordinates $(s,u)$. Let us set
\begin{eqnarray}
s = \log \left(-{U\over V}\right)\,,\qquad u=-\log(-U V)\,,\qquad
U\in (0, +\infty)\,,\quad  V \in(-\infty, 0)\,;
\label{coor_su}
\end{eqnarray}
the transformation $(U,V)\mapsto(s,u)$ is one to one between the sets
$(0, +\infty)\times(-\infty, 0)$
and $\Real^2$.
By compositions with \eqref{coordsu_dS} we obtain the transformation
\begin{equation}
{t}={\ell\over 2(1-B^2)}\log\left(-{ U\over V}\right)~, \quad
x=\ell \, {1+UV\over 1-UV}~,
\label{xuv}
\end{equation}
which is a diffeomorphism between $(0, +\infty)\times(-\infty, 0)$
and the inner region
$I$.
The first cosmological horizon ${x}=-\ell$ corresponds
to $U\to +\infty$ or  $V\to -\infty$, while the second cosmological horizon ${x}=\ell$ coincides with $UV=0$.
Now the metric \eqref{gsu} reads
\begin{equation}
\label{Eq:g_UV}
{\bf g} = \frac{1}{k^2 (1 - U V)^2}
\left[ -4 d U d V + \bigg(  B^2 (1 - U V)^2 + (1 - B^2) (1 + U V)^2 \bigg)
(d \vartheta^2 + \sin^2 \vartheta\, d \varphi^2) \right]\,.
\end{equation}
It is evident that this metric is regular on the cone $U V=0$ and can be extended beyond the corresponding horizon to the region
\begin{equation}
{\mathscr{R}} := \{ (U,V)\in\Real^2~|~\, U V < 1 \}\,,
\label{Region_R}
\end{equation}
which is bounded by the two branches of the hyperbola $U V = 1$, corresponding to the spacelike infinity ${x} = +\infty$. The two branches of the hyperbola $U V = -1$ correspond to the throat $x = 0$. To go on, let us extend the transformation \eqref{xuv} setting
\begin{equation}
{t}={\ell\over 2(1-B^2)}\log\left|{ U\over V}\right|~, \quad
x=\ell \, {1+UV\over 1-UV}
\label{xuvest}
\end{equation}
whenever this makes sense.
The map
$(U,V) \mapsto x$ is smooth throughout the region ${\mathscr{R}}$, while $(U,V) \mapsto t$ is well defined
and smooth on the subregion $\{ (U,V) \in {\mathscr{R}}~|~U V \neq 0 \}$. The correspondence
$(U,V) \mapsto (t,x)$ gives diffeomorphisms
between the following pairs of regions: $( 0 , +\infty)\times( -\infty , 0)$ and $I$ (as already shown);
$( -\infty,0)\times( 0,+\infty )$ and
$I$;
$\{ (U,V)\in(0,+\infty)^2\,|\, U V < 1 \}$ and the exterior region $E^+$;
$\{(U,V)\in(-\infty,0)^2\,|\, U V < 1 \}$ and the exterior region $E^+$.
Under each one of these four diffeomorphisms, the metric of
Eq.~\eqref{Eq:dSWH1} takes the form \eqref{Eq:g_UV}. To conclude we note that, writing
$\Phi$ as in Eq.~\eqref{Eq:StaticSolutionsPotdS} and $x$ as in Eq.~\eqref{xuvest} we obtain a smooth extension of the scalar field $\Phi$ to the whole region ${\mathscr{R}}$.

\subsection{Extending spacetime further}
\label{Subsubsection:extension_ds}

\begin{figure}[htbp]
    \begin{centering}
        \includegraphics[scale=0.6]{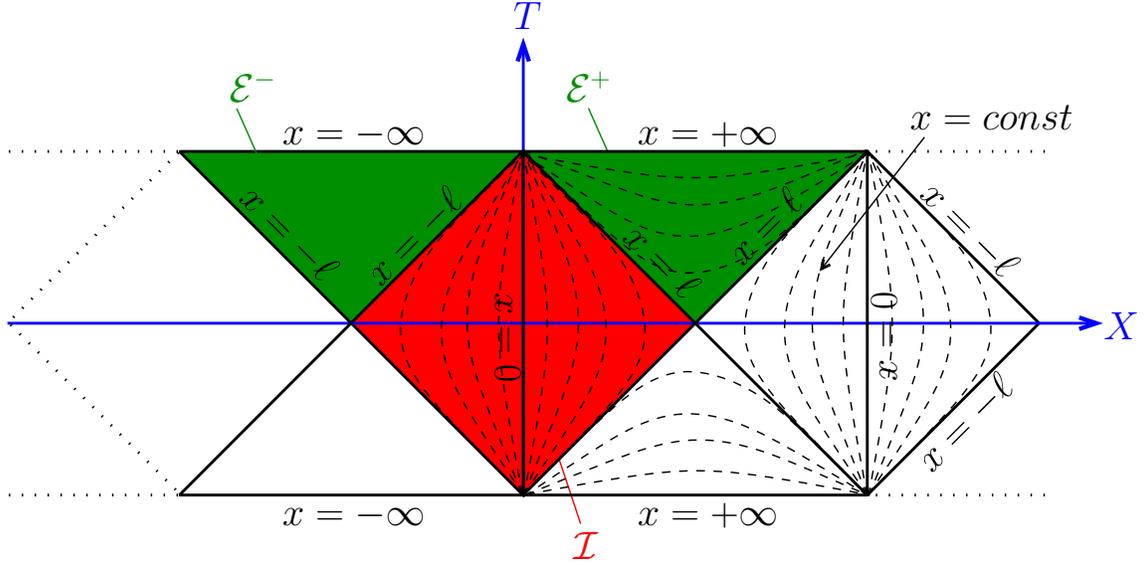}
        \par\end{centering}
   \caption{
Penrose diagram showing the strip $\mathcal{S}$ in the final extended spacetime of our dS wormhole (Eqs.~(\ref{S},\ref{g_T_X})). The dashed lines are lines with constant $x$, determined according to Eq.~\eqref{x_T_X}. Also indicated are the red diamond region $\mathcal{I}$ and the green triangular regions $\mathcal{ E}^{\pm}$ of Eq.~\eqref{I_E} which correspond to the original regions~\eqref{region_I_E} in the $(t,x)$ coordinate space; the same can be said of the images of $\mathcal{I}$ and $\mathcal{ E}^\pm$ under any translation $\mathfrak{T}^{2 h} : (T,X)\mapsto(T,X+2h\pi)$ ($h\in \mathbb{Z}$). Applying the time reflection $\mathfrak{R}: (T,X)\mapsto(-T,X)$ to the triangles $\mathcal{E}^\mp$ and to the translated triangles mentioned before, one obtains other regions which are isometric to $E^\mp$.}
\label{Fig:dSWH}
\end{figure}

We now consider a ``compactification'' of the extended region ${\mathscr{R}}$ \eqref{Region_R} based on the reparametrization
\begin{equation}
U=\tan \uu\,,\qquad V=\tan \vv\,.
\label{uhat_vhat}
\end{equation}
We know that the cone $UV=0$ and the limits $U\to +\infty$, $V\to -\infty$ and $U\to -\infty$, $V\to +\infty$ correspond to the horizons $x=\pm\ell$ in \eqref{xuvest}; according to Eq.~\eqref{uhat_vhat} the cone and the indicated limits are associated with finite values of $\uu$ and $\vv$, so the effect of the above transformation is to bring both the horizons to finite distances. One could use $\uu$ and $\vv$ as an alternative set of coordinates and reexpress the metric~\eqref{Eq:g_UV} and so on; but the situation can be described in a simpler way by making a further transformation (essentially, a rotation of ${\pi\over 4}$ and a translation of the axes)
\begin{equation}
\uu ={T\over 2} - {X\over 2} + {\pi\over 4}\,,\qquad
\vv = {T\over 2} + {X\over 2} - {\pi\over 4} \,.
\label{T_X}
\end{equation}
The composition of Eqs.~(\ref{uhat_vhat},\ref{T_X}), whenever they make sense, gives
\begin{equation}
U = \tan \left({T\over 2} - {X\over 2} + {\pi\over 4}\right)\,,\qquad
V = \tan \left({T\over 2} + {X\over 2} - {\pi\over 4}\right)\,;
\label{U_V_T_X}
\end{equation}
the application $(T,X)\mapsto(U,V)$ is a bijection between the regions ${\cal R}$ and  $\mathscr{R}$, where $\mathscr{R}$ is defined by Eq.~\eqref{Region_R}  and
 $\mathcal{R}:= \{ (T,X)\in\Real^2 | -{\pi\over 2}< T <{\pi\over 2},\, -{\pi\over 2} < X-T,X + T<{3\over 2}\pi\}$ .
In the coordinates $(T,X, \vartheta,\varphi)$ the metric~\eqref{Eq:g_UV} assumes the form
\begin{equation}
{\bf g} = \frac{1}{k^2\cos^2 T}\bigg[ -dT^2 + dX^2
+ \bigg( {B^2}\cos^2 T +( 1-B^2) \sin^2 X \bigg) (d \vartheta^2 + \sin^2 \vartheta\, d \varphi^2) \bigg],
\label{g_T_X}
\end{equation}
which clearly admits a further extension to the region $\mathcal{ S} \times S^2$,
where we have defined
\begin{equation}
 \mathcal{ S} := \left\{ (T,X)\in\mathbb{R}^2\,|\, -\frac{\pi}{2}<T<\frac{\pi}{2} \right\}\,.
\label{S}
\end{equation}
Equations~(\ref{g_T_X},\ref{S}) provide the final form of our dS wormhole spacetime; the strip $\mathcal{ S} $ is represented in Fig.~\ref{Fig:dSWH}, which also accounts for some facts illustrated hereafter. Note that the metric  (\ref{g_T_X}) is invariant under the spatial translation, the spatial reflection and the time reflection
\begin{equation}
\mathfrak{T} : (T,X) \mapsto (T, X + \pi)\,,\qquad\mathfrak{S} : (T, X) \mapsto (T, \pi - X)\,,\qquad \mathfrak{R} : (T,X) \mapsto (-T,X)\,.
\label{transf}
\end{equation}
Let us also remark that, in the limit case $B\to 0$, the expression (\ref{g_T_X}) reduces to the familiar representation of the dS metric as a conformal factor times the line element of the static Einstein universe.
For any $B>0$, the connection between the spacetime (\ref{g_T_X},\ref{S}) and the original setting (\ref{Eq:dSWH1},\ref{region_I_E})
is understood by expressing the original variables $(t,x)$ in terms of the new variables $(T,X)$. To this purpose we note that the composition of the transformations (\ref{xuvest},\ref{U_V_T_X}), whenever they make sense, gives
\begin{equation}
t = {\ell\over 2(1-B^2)}\log\left| {\sin T + \cos X \over \sin T - \cos X} \right|\,,\qquad
x = \ell {\sin X\over \cos T}\,.
\label{x_T_X}
\end{equation}
The map of $(T,X)\mapsto x$ is everywhere smooth on $\mathcal{ S} $, while the map of $(T,X)\mapsto t$ has singularities at the points of $\mathcal{S}$ where the argument of the logarithm vanishes or diverges; this occurs at points where $\sin T =\mp \cos X$, which are just the points where $x=\mp \ell$. Moreover, we note that
$(t,x) \circ \mathfrak{T} = (-t ,-x)$,
$(t,x) \circ \mathfrak{S} = (-t, x)$ and
$(t,x) \circ \mathfrak{R} = (-t,x)$; the behavior of $\mathbf{g},t,x$ under $\mathfrak{T}$ implies the invariance of each one of these three objects under the translation $\mathfrak{T}^2:(T,X)\mapsto(T,X+ 2\pi)$.

To go on, let us now introduce the diamond $\mathcal{I}$ and the triangles $\mathcal{ E}^\mp$ defined by
\begin{align}
\nonumber
\mathcal{I} &:=\left\{(T,X)\in\Real^2\,|\, - {\pi\over2} < T-X,T+X < {\pi\over2}\right \}\,,\\
\label{I_E}
\mathcal{ E}^- &:=\left\{(T,X)\in\Real^2\,|\, T < {\pi\over2},\, T-X > {\pi\over 2},\,T+X > -{\pi\over 2}\right \}\,,
\\
\mathcal{ E}^+ &:=\left\{(T,X)\in\Real^2\,|\, T<{\pi\over2},\,T-X > -{\pi\over 2},\, T+ X > {\pi\over 2}\right \}
\nonumber
\end{align}
(see again Fig.~\ref{Fig:dSWH}); then the map $(T,X)\to(t,x)$, described by Eq.~\eqref{x_T_X}, gives isometric diffeomorphisms between $\mathcal{I}$ and $I$, between $\mathcal{ E}^-$ and $E^-$, and between $\mathcal{ E}^+$ and $E^+$, where $I$ and $E^\mp$ are, respectively, the internal region and the two exterior regions~(\ref{region_I_E}) with the metric \eqref{Eq:dSWH1}.
Moreover, we have that $x=\pm\ell$ along the sides of $\mathcal{ I}$, $x=-\ell$ and $x=-\infty$ along the sides of $\mathcal{ E}^-$ and $x=\ell$ and $x=+\infty$ along the sides of $\mathcal{ E}^+$ (see once more Fig.~\ref{Fig:dSWH}). It is easy to construct infinitely many replicas of the previous statement using the previous information of the behavior of $\mathbf{g},t,x$ under the transformations~(\ref{transf}). For example, using the fact that $\mathbf{g},t,x$ are invariant under all the iterates $\mathfrak{T}^{2 h}: (T,X)\mapsto(T,X+ 2h\pi)$ ($h\in\mathbb{Z}$), one can readily show that for each $h\in\mathbb{Z}$, the map~\eqref{x_T_X} gives isometric diffeomorphisms between
 $\mathfrak{T}^{2 h}(\mathcal{I})$ and $I$, between $\mathfrak{T}^{2 h}(\mathcal{ E}^-)$ and $E^-$, and between $\mathfrak{T}^{2 h}(\mathcal{E}^+)$ and $E^+$. Moreover, by applying the time reflection $\mathfrak{R}$ to each one of the translated triangles $\mathfrak{T}^{2 h}(\mathcal{E}^\mp)$ one gets other regions isometrically diffeomorphic to $E^\mp$.

Finally, let us recall that we have already noted that the points $(T,X)$ where  Eq.~\eqref{x_T_X} gives singularities for $t$ are just the points at which the same equation gives $x=\pm\ell$; so from the viewpoint of the extended manifold $\mathcal{S}\times S^2$, the apparent singularities at $x=\pm\ell$ of the original metric \eqref{Eq:dSWH1} are just due to the singularities of $t$ as a coordinate on $\mathcal{S}$.

Up to now, we have not considered the scalar field $\Phi$. The prescription
\begin{equation}
 \Phi = \sqrt{\frac{2}{\kappa}}\arctan\frac{x}{b}\,,\qquad \mbox{ with $x$ as in Eq. \eqref{x_T_X}}
 \label{phi}
 \end{equation}
 gives a smooth function everywhere on $\mathcal{ S}$, with the properties
 $\Phi\circ\mathfrak{T}=-\Phi$, $\Phi\circ\mathfrak{T}^{2}=\Phi$ and so on.
 The triple $\mathcal{S}\times S^2,\mathbf g,\Phi$ in Eqs.~(\ref{S},\ref{g_T_X},\ref{phi})
 is a solution to the Einstein-scalar equations (with field self-potential
 $V(\Phi)$ as in \eqref{Eq:StaticSolutionsPotdS}) .

Of course, the extended spacetime $\mathcal{ S}\times S^2$ has the topology of
$\mathbb{R}^2\times S^2$. For any fixed $p=1,2,3,...$ we can take the quotient
of the strip $\mathcal{ S}$ with respect to the iterated translation $\mathfrak{T}^{p}$;
the quotient $\mathcal{ S}/\mathfrak{T}^{p}$ has the topology of $\Real\times S^1$
and the metric \eqref{g_T_X} can be projected on $(\mathcal{ S}/\mathfrak{T}^{p})\times S^2$,
thus getting a new spacetime with the topology $\Real\times S^1\times S^2$. The function $\Phi$ of Eq. \eqref{phi} is projectable on this quotient
spacetime for $p$ even, since in this case $\Phi\circ\mathfrak{T}^p=\Phi$;
on the contrary, $\Phi$ is not projectable for $p$ odd because
$\Phi\circ\mathfrak{T}^p=-\Phi$. Finally, let us mention that
all spacetimes $\mathcal{ S} \times S^2$ and
$(\mathcal{ S}/\mathfrak{T}^{p}) \times S^2$ ($p=1,2,3,\ldots$) are time
orientable: in fact, $\partial/\partial T$ is a smooth timelike
vector field, defined everywhere on $\mathcal{ S} \times S^2$
and projectable on $(\mathcal{ S}/\mathfrak{T}^{p}) \times S^2$
both for $p$ even and for $p$ odd. One could also consider the quotients $(\mathcal{ S}/(\mathfrak{T}^{p}\circ \mathfrak{R}))$ with $p=1,2,3,\ldots$ involving the time reflection, which yield smooth spacetimes which are, however, not time orientable.

\subsection{Linear instability of the dS wormhole in the inner region}
\label{Subsubsection:intstab_ds}

If one confines the attention to the spacetime  $(I\times S^2, {\bf g}, \Phi)$,
where $I$ is the inner region \eqref{region_I_E} in $(t,x)$ space and $\Phi, {\bf g}$ are as in Eqs.~(\ref{Eq:StaticSolutionsPotdS},\ref{Eq:dSWH1}), the analysis of linearized perturbations for the Einstein-scalar equations is rather simple in the framework of this paper.

First of all, one replaces the coordinates $(t,x)$ with the coordinates $(s,u) \in \Real^2$ defined by Eq.~\eqref{coordsu_dS}. After this, one should in principle apply the general scheme of sections \ref{Sec:LinearPerturbation}-\ref{Sec:Decoupling} (in the coordinates $(s,u)$) to the linearized perturbations of this solution, ultimately yielding a master equation. As a matter of fact, it is not even necessary to carry on this construction and it suffices to use the following trick: since the dS wormhole under analysis is connected to the AdS wormhole of sections~\ref{Subsection:AdS},\ref{SubSec:Pertads} through the formal replacement rules $(k,B,s,u) \mapsto (i k, i B, i s, i u)$ (see the comments after Eqs.~(\ref{Eq:dSWH1},\ref{gsu})), the master equation for the perturbed dS wormhole can be obtained, making formally the same replacements in Eqs.~(\ref{Eq:MasterAdS},\ref{Eq:PotentialB}) of the AdS case. In conclusion, the master equation governing linear perturbations of the dS wormhole, in an unknown function $\chi(s,u)$, reads
\begin{equation}
\left[ \frac{\partial^2}{\partial s^2}
  - \frac{\partial^2}{\partial u^2}  +
 {\cal {V}} \right] \chi = 0\,,
\end{equation}
and involves the potential
\begin{equation}
{\mathcal{V}}(u) \equiv {\mathcal{V}}_{B}(u) :=
- \frac{B^2 \left(2-B^2+\cosh u\right) }{\left(-1+2 B^2 + \cosh u\right)^2}
\end{equation}
($s,u \in \Real$); note that ${\mathcal{V}}(u)$ is everywhere negative and
vanishes like $-1/\cosh u$ for $u \to \pm \infty$.
The corresponding Schr\"odinger operator
$$
H := - \frac{d^2}{d u^2} + \mathcal{V}
$$
is selfadjoint in $L^2(\Real, du)$, and can be
analyzed by standard methods, including Sturm oscillation theory
(\footnote{
Application of Sturm theory relies on
the zero energy Schr\"odinger equation $[-d^2/d u^2 + \mathcal{V} ]\chi_0 = 0$.
The general solution of this equation is obtained
from the analogous solution \eqref{Eq:AdSZeroMode} for the AdS case
with the formal replacements $(u,B) \mapsto (i u, i B)$ and reads
$$
\chi_0(u) = C_1 {\sinh {u \over 2} \over \sqrt{-1+2 B^2 + \cosh u}}
 + C_2 {2 u \sinh{u \over 2} - 4 B^2 \cosh{u \over 2}\over \sqrt{-1+2 B^2 + \cosh u}}\,
$$
($C_1, C_2 \in \Complex$). One has
$\chi_0 \in L^2(\Real, d u)$ if and
only if $C_1=C_2=0$, thus zero is not an eigenvalue of $H$. If
$C_1 \in \Real \setminus \{0 \}$ and $C_2=0$ it is evident that $\chi_0$ has a unique
zero in $\Real$ (namely, $u=0$). If $C_2 \in \Real \setminus \{0 \}$ and $C_1 \in \Real$,
one can show that $\chi_0$ possesses two
zeroes in $\Real$ (via an analysis rather similar
to that given for the function $\chi_0$ of the AdS case \eqref{Eq:AdSZeroMode};
see, in particular, the footnote which accompanies this equation).
Summing up, the \emph{minimal} number of zeroes of the real, non identically vanishing
solutions $\chi_0$ of the zero energy equation is \emph{one}.
The Sturm oscillation theorem (see Theorem 14.8 of \cite{Weid-Book}) states that such a minimal number
of zeroes is the number of negative eigenvalues of $H$.
So, $H$ has a unique negative eigenvalue; in addition, due
to general facts on Schr\"odinger operators (and to
the previous remark that $0$ is not an eigenvalue), $H$ has
continuous spectrum $[0,+\infty)$.}).

In this way, the spectrum
of $H$ is found to consist of a unique negative eigenvalue and of the continuous
spectrum $[0,+\infty)$. The situation is similar to that of the reflection-symmmetric
Ellis-Bronnikov wormhole: the system is linearly unstable, and the general
solution of the master equation has the form given by Eq.~\eqref{solbro} (with the
variables $(t,x) \in \Real^2$ appearing therein replaced by the present variables $(s,u) \in \Real^2$).

\subsection{Linear instability of the extended dS wormhole?}

For a full understanding of the subject under discussion,
linearized perturbations of the Einstein-scalar equations
should be treated on the extended spacetime $\mathcal{S}\times S^2$
of subsection \ref{Subsubsection:extension_ds}
(or on the quotients $(\mathcal{S}/ \mathfrak{T}^p)\times S^2$),
possibly in a gauge-invariant fashion.
The discussion of this problem would bring us outside the scope of
the present paper, since the extended spacetime $\mathcal{S}\times S^2$ is not static.
One can reasonably expect that the instability result of the previous subsection
about the inner region will
eventually produce a precise statement of linear instability for the extended dS wormhole.
However, we prefer to postpone these matters to future works; let us also mention
that the notion of linear instability is not so obvious if one perturbs
a nonstatic spacetime, and requires in our opinion a general discussion
before reconsidering the specific case of the extended dS wormhole.

\section{Conclusions}
\label{Sec:Conclusions}

In this work we have analyzed the linear stability of a class of static, spherically symmetric wormhole solutions in GR minimally coupled with a self-interacting phantom scalar field. To this purpose, we have provided a gauge-invariant perturbation formalism that describes the dynamics of linearized, spherical but time-dependent perturbations of the metric and of the scalar field, resulting in a coupled $2\times 2$ linear wave system subject to a constraint (see Eqs.~(\ref{Eq:WaveSystem},\ref{Eq:WaveConstraint})). Provided that a nontrivial, time-independent solution is known (as is usually the case when a family of static solutions is known) we have shown that this system can be decoupled to yield a master wave equation which is manifestly gauge-invariant and regular at the throat. This construction relies on a basic requirement (of course satisfied by the examples that we treat): the derivative $\Phi'$ of the (background) scalar field should vanish nowhere. The relevance of this condition in our approach is indicated by the almost ubiquitous presence of the reciprocal $1/\Phi'$ in the equations of sections~\ref{Sec:LinearPerturbation}-\ref{Sec:Decoupling}.

Based on our formalism we have rederived the regular master equation first obtained in~\cite{jGfGoS09a}, describing linear spherical perturbations of the Ellis-Bronnikov wormhole in a fully gauge-invariant setting and without intermediate steps involving singularities at the throat. (For an alternative approach which treats the reflection-symmetric case in a fixed gauge, see~\cite{fCfPlP19}.) Furthermore, we have analyzed the linear stability of an AdS wormhole introduced in~\cite{kBjF06}, for which the scalar field is subject to a nontrivial self-interaction term, and we have shown that this solution is linearly unstable as well. In both examples, the instability is characterized by a unique mode growing exponentially in time, associated with a bound state of negative energy of the Schr\"odinger operator arising in the master equation. As discussed in section~\ref{Sec:SpectralD} the associated instability times are rather short (of the order of a light-crossing time corresponding to the areal radius of the throat.)

Based on spectral analysis, we have also provided a detailed and rigorous discussion for the mode decomposition of the solutions to the master wave equations in both the aforementioned examples, which revealed that besides the modes growing exponentially in time, there might also be linearly growing modes, while all the remaining modes are oscillatory. In particular, the AdS wormhole has infinitely many normalizable, oscillatory modes in addition to the pair of exponentially growing and decaying modes associated with the unique bound state of negative energy of the Schr\"odinger operator.

In the last section, which is admittedly outside the mainstream of the present paper, we have also sketched the discussion of a dS wormhole with horizons, whose spacetime has a natural nonstatic extension; in this case we have provided a linear instability result, which, however, refers only to the static spacetime region within the horizons.

Let us conclude with some remarks on the possible future developments of the present work. We have already mentioned that the linear stability theory for nonstatic wormhole solutions, and its application to the (extended) dS wormhole, deserves further work in our opinion. Sticking to the case of static wormholes and of their linearized perturbations, we think that the forthcoming issues (i-ii) are worthy of future investigation:

(i) A basic requirement of our approach, recalled above, is the condition that $\Phi$ have no critical points. Removing this requirement would be interesting since, recently, a large class of new wormhole solutions of the Einstein-scalar equations has been found~\cite{Carvente:2019gkd}, generalizing previous work~\cite{Dzhunushaliev:2017syc}, in which the scalar field $\Phi$ has an extremum at the throat. Since $r$ has a global minimum at the throat and $r'$ converges to zero as fast as or faster than $\Phi'$, it turns out that the gauge-invariant quantity $C$ defined in Eq.~(\ref{Eq:CGI}) is still well defined; unfortunately, it is unclear if a decoupled equation for $C$ can be obtained that is regular at the throat. In connection with this problem, one could try to recover the S-deformation method of~\cite{jGfGoS09a,kBjFaZ11} (see the discussion in the Introduction; the formulation of this method in~\cite{kBjFaZ11} indeed considers the gauge-invariant quantity $C$). However, when the potential $V(\Phi)$ is nonzero, this method seems to require the numerical integration of a Riccati-type equation to find the regularized potential, and further one still needs to justify \emph{a posteriori} the validity of the transformed equation at the throat. An alternative possibility consists in applying a variation of the approach discussed in this article, in which $\Phi'$ is absent from all denominators, thanks to the use of new gauge-invariant quantities in place of the functions $A,C,E$ of Eqs.~(\ref{Eq:AGI}-\ref{Eq:EGI}); at present, it is not clear to us whether this will be possible.

(ii)  Let us propose the following question: is there a \emph{deep} geometrical reason for which our present approach succeeds, in certain cases, in decoupling the perturbation equations~(\ref{Eq:WaveSystem}-\ref{Eq:WaveConstraint}) and reducing them to a single, scalar master equation? Typically, the possibility of reducing to a simpler form a PDE or of a system of PDEs is due to the presence of a Lie group of symmetries; an interpretation of this kind could perhaps be given for our decoupling method. As already recalled, our approach uses a static solution of Eqs.~(\ref{Eq:WaveSystem}-\ref{Eq:WaveConstraint}), arising from variations with respect to the parameters of a \emph{family} of static wormhole solutions. The availability of such parametric families could perhaps be interpreted in terms of a Lie group of symmetries, acting on the static solutions of the Einstein-scalar system; if so, it would be interesting to understand the interplay of these symmetries with the linearized perturbation equations.

\acknowledgments \noindent F.C. and L.P. were supported by: INdAM,
Gruppo Nazionale per la Fisica Matematica; Universit\`{a} degli
Studi di Milano. LP was also supported by: INFN; MIUR, PRIN 2010
Research Project \virg Geometric and analytic theory of
Hamiltonian systems in finite and infinite dimensions." O.S. was
partially supported by a CIC grant to Universidad Michoacana. We thank an anonymous referee for pointing out to us Ref.~\cite{Torii} and for the suggestion to extend the stability
analysis to the dS case.

\appendix
\section{On the master equation for the reflection-symmetric Ellis-Bronnikov wormhole}
\label{appebro}

We refer to the master equation for this wormhole in the formulation
(\ref{masterbro}), based on the Hilbert space $\Hilb := L^2(\Real, dx)$
of Eq.~(\ref{hilbro}) and on the selfadjoint operator $\Hop$
of Eq.~(\ref{hbro}), of domain $\mathfrak{D} \subset \Hilb$. We keep
all notations introduced after these equations; in particular
$\langle~|~\rangle$ and $\|~\|$ are the natural inner product and
norm of $\Hilb$.
\\\\
\textbf{Relevant facts on the operator $\Hop$ and its
spectral features.}
In the discussion following Eqs.~(\ref{masterbro}-\ref{hbro}) we have mentioned
a system made by a normalized eigenfunction $e_1$ for the unique eigenvalue
$\mu_1 < 0$ of $\Hop$, and by a pair of improper eigenfunctions
$e_{i \lambda}$ ($i=1,2$) (lying in $C^\infty(\Real)$ but not in $\Hilb$)
for each $\lambda>0$. We have called this system
a generalized orthonormal basis of $\Hilb$, which means that the following
conditions hold \cite{Ber}:
\begin{itemize}
\setlength\itemsep{-0.1cm}
\item[(a)] Consider the space $C_c(\Real) \equiv \mathfrak{C} \subset \Hilb$, made of
the continuous functions $f : \Real \to \Complex$ with compact
support. For $f \in \mathfrak{C}$, consider the usual inner product
$\langle e_1 | f \rangle = \int_{\Real} d x \bar{e}_1(x) f(x)$ and
define in addition a ``generalized inner product''
$\langle e_{i \lambda} | f \rangle := \int_{\Real} d x \, \bar{e}_{i \lambda}(x) f(x)$
(the integral converges since $\bar{e}_{i \lambda} f \in \mathfrak{C}$); then, the maps
\begin{equation} (0,+\infty) \ni \lambda \mapsto \langle e_{i \lambda} | f \rangle \in
\Complex \quad(i=1,2)
\end{equation}
are both in $L^2((0,+\infty), d \lambda$).
\item[(b)] For $i=1,2$, the linear map $\mathfrak{C} \subset \Hilb \to L^2((0,+\infty), d \lambda)$,
$f \mapsto \Big( \lambda \mapsto \langle e_{i \lambda} | f \rangle \Big)$
is continuous with respect to the norms of the Hilbert spaces $\Hilb$ and $L^2((0,+\infty), d \lambda)$;
thus, by the density of  $\mathfrak{C}$ in $\Hilb$, this map has a unique continuous (and linear)
extension to $\Hilb$, that we write as
\begin{equation}
\Hilb \to L^2((0,+\infty), d \lambda)\,, \quad f \mapsto
\Big( \lambda \mapsto \langle e_{i \lambda} | f \rangle \Big)\,.
\end{equation}
For each $f \in \Hilb$, the map $\lambda \mapsto \langle e_{i \lambda} | f \rangle$
is said to give the ``generalized inner products'' between the $e_{i \lambda}$'s and $f$.
\item[(c)] Consider the direct sum
Hilbert space $\Complex \oplus L^2((0,+\infty), d \lambda)
\oplus L^2((0,+\infty), d \lambda)$ with its natural inner product; then the linear map
\begin{equation}
\Hilb \to \Complex \oplus L^2((0,+\infty), d \lambda)
\oplus L^2((0,+\infty), d \lambda)\,, \quad
f \mapsto \Big(\langle e_1 | f \rangle, \, \lambda \mapsto \langle e_{1 \lambda} | f \rangle,
\, \lambda \mapsto \langle e_{2 \lambda} | f \rangle \Big)
\label{unit}
\end{equation}
is a unitary, i.e., it is one-to-one and preserves inner products:
\begin{equation}
\langle f | \ell \rangle = \overline{\langle e_1 | f \rangle} \langle e_1 | \ell \rangle
+ \sum_{i=1}^2 \int_{0}^{+\infty} \hspace{-0.4cm} d \lambda \, \overline{\langle e_{i \lambda} | f \rangle}
\langle e_{i \lambda} | \ell \rangle \quad (f,\ell \in \Hilb)\,.
\end{equation}
\end{itemize}
\noindent
The forthcoming items describe some consequences of conditions (a-c):
\begin{itemize}
\setlength\itemsep{-0.1cm}
\item[(i)] For $F \in L^2((0,+\infty), d \lambda)$ and $i = 1,2$ there is a unique element of $\Hilb$,
indicated with $\int_{0}^{+\infty} d \lambda F(\lambda) e_{i \lambda}$, such that
\begin{equation}
\label{Eq:wi}
\left \langle \int_{0}^{+\infty} \hspace{-0.4cm} d \lambda F(\lambda) e_{i \lambda} \bigg| \ell \right \rangle =
\int_{0}^{+\infty} \hspace{-0.4cm} d \lambda \overline{F}(\lambda) \langle e_{i \lambda} | \ell \rangle
\quad \mbox{for all $\ell \in \Hilb$}
\end{equation}
(note that the integral on the right-hand side of Eq.~(\ref{Eq:wi}) exists, involving
the product of two functions which are both in $L^2((0,+\infty), d \lambda)$).
The element $\int_{0}^{+\infty} d \lambda F(\lambda) e_{i \lambda} \in \Hilb$ is called
the weak integral of the function $\lambda \mapsto F(\lambda) e_{i \lambda}$.
\item[(ii)] The inverse of the unitary map~(\ref{unit}) can be expressed in terms of weak integrals; more
precisely, such inverse is the map
\begin{equation}
\Complex \oplus L^2((0,+\infty), d \lambda)
\oplus L^2((0,+\infty), d \lambda) \to \Hilb\,,\quad
(\alpha, F_1, F_2) \mapsto \alpha e_1 + \sum_{i=1}^2 \int_{0}^{+\infty} \hspace{-0.4cm} d \lambda F_i(\lambda) e_{i \lambda}\,.
\label{unitinv}
\end{equation}
\end{itemize}
\noindent
The fact that the composition of the maps~(\ref{unit},\ref{unitinv}) is the identity map
$\Hilb \to \Hilb$, $f \mapsto f$ can be written explicitly as follows: for each $f \in \Hilb$,
\begin{equation}
f = \langle e_1 | f \rangle e_1 + \sum_{i=1}^2 \int_{0}^{+\infty}
\hspace{-0.4cm} d \lambda \langle e_{i \lambda} | f \rangle e_{i \lambda}\,.
\label{frep}
\end{equation}
The identity~(\ref{frep}) is said to give the expansion of $f$ in terms of the generalized orthonormal basis under consideration.

Up to now, we have not used the fact that $e_1$ is an eigenfunction
of $\Hop$ with eigenvalue $\mu_1 < 0$, nor the fact that $e_{i \lambda}$
is an ``improper eigenfunction'' with ``eigenvalue'' $\lambda$ for
$i=1,2$ and all $\lambda > 0$. These facts yield the following
representation for the operator $\Hop$ and its domain $\mathfrak{D}$:
\begin{equation}
\begin{gathered}
\mathfrak{D} = \{ f \in \Hilb~|~\big(\lambda \mapsto \lambda \langle e_{i \lambda}| f \rangle \big)
\in L^2((0,+\infty), d \lambda)~\mbox{for $i=1,2$} \}\,. \\
\mbox{For}~ f \in \mathfrak{D}: \quad
\langle e_1 | \Hop f \rangle =
    \mu_1\langle e_1 | f \rangle\,,
~\langle e_{i \lambda} | \Hop f \rangle =
    \lambda\langle e_{i \lambda} | f \rangle\,,
~~\mbox{i.e.}\,,
~~\Hop f = \mu_1 \langle e_1 | f \rangle e_1 +
    \sum_{i=1}^2 \int_{0}^{+\infty}
    \hspace{-0.4cm} d \lambda \, \lambda \langle e_{i \lambda} | f \rangle e_{i \lambda}\,.
\end{gathered}
\label{reph}
\end{equation}
As well known, a functional calculus exists for selfadjoint Hilbert space
operators (see, e.g., \cite{Schmudgen-Book}). This allows to define an operator
$\mathscr{F}(\Hop) : \mathfrak{D}^{\mathscr{F}} \subset \Hilb \to \Hilb$
for each (Borel-) measurable function $\mathscr{F} : \sigma(\Hop) \to \Complex$,
where $\sigma(\Hop) = \{ \mu_1 \} \cup [0,+\infty)$ is the spectrum
of $\Hop$ and $\mathfrak{D}^{\mathscr{F}}$ is a suitable
domain, determined by ($\Hop$ and) $\mathscr{F}$; the operator $\mathscr{F}(\Hop)$ is selfadjoint if
$\mathscr{F}$ is real valued. Making reference to the
previously mentioned, generalized orthonormal basis
of eigenfunctions of $\Hop$, one can prove the following statements:
$e_1 \in \mathfrak{D}^{\mathscr{F}}$ and $\mathscr{F}(\Hop) e_1 = \mathscr{F}(\mu_1) e_1$;
$\mathfrak{D}^{\mathscr{F}} =
\{ f \in \Hilb~|~\big(\lambda \mapsto \mathscr{F}(\lambda) \langle e_{i \lambda}| f \rangle \big)
\in L^2((0,+\infty), d \lambda)~\mbox{for $i=1,2$} \}$; for all
$f \in \mathfrak{D}^{\mathscr{F}}$, one has
$\langle e_1 | \mathscr{F}(\Hop) f \rangle = \mathscr{F}(\mu_1)
\langle e_1 | f \rangle$ and
$\langle e_{i \lambda} | \mathscr{F}(\Hop) f \rangle = \mathscr{F}(\lambda)
\langle e_{i \lambda} | f \rangle$. For our purposes it is
important to consider the choice $\mathscr{F}(\gamma) := | \gamma|^{1/2}$ for
all $\gamma \in \sigma(\Hop)$, producing a selfadjoint operator that we
indicate with
\begin{equation}
| \Hop |^{1/2} : \mathfrak{D}^{1/2} \subset \Hilb \to \Hilb
\end{equation}
and that behaves as follows in relation to our generalized orthonormal
basis:
\begin{equation}
e_1 \in \mathfrak{D}^{1/2}\,,\quad|\Hop|^{1/2} e_1 = |\mu_1|^{1/2} e_1\,.
\end{equation}
\begin{equation}
\begin{gathered}
\mathfrak{D}^{1/2} = \{ f \in \Hilb~|~\big(\lambda \mapsto \lambda^{1/2} \langle e_{i \lambda}| f \rangle \big)
\in L^2((0,+\infty), d \lambda)~\mbox{for $i=1,2$} \}\,.~\mbox{For}~ f \in \mathfrak{D}^{1/2}:\\
{~} \hspace{-0.5cm}
\langle e_1 \big| |\Hop|^{1/2} f \rangle = |\mu_1|^{1/2} \langle e_1 | f \rangle,
\langle e_{i \lambda} \big| |\Hop|^{1/2} f \rangle = \lambda^{1/2}
\langle e_{i \lambda} | f \rangle,~\mbox{i.e.},~
|\Hop|^{1/2} f = |\mu_1|^{1/2} \langle e_1 | f \rangle e_1 + \sum_{i=1}^2 \int_{0}^{+\infty}
\hspace{-0.6cm} d \lambda \, \lambda^{1/2} \langle e_{i \lambda} | f \rangle e_{i \lambda}\,.
\end{gathered}
\label{reph12}
\end{equation}
Finally, let us make explicit the Hilbert space structures for $\mathfrak{D}$ and $\mathfrak{D}^{1/2}$
mentioned before Eq.~(\ref{solbro}); these are provided by the (complete) inner products
$\langle~|~\rangle_{\mathfrak{D}} : \mathfrak{D} \times \mathfrak{D} \to \Complex$ and
$\langle~|~\rangle_{\mathfrak{D}^{1/2}} : \mathfrak{D}^{1/2} \times \mathfrak{D}^{1/2} \to \Complex$, where
\begin{equation}
\langle f | \ell \rangle_{\mathfrak{D}} :=
\langle f | \ell \rangle + \langle \Hop f | \Hop \ell \rangle
= (1 + \mu_1^2) \, \overline{\langle e_1 | f \rangle} \langle e_1 | \ell \rangle
+ \sum_{i=1}^2 \int_{0}^{+\infty} \hspace{-0.4cm} d \lambda \,
(1 + \lambda^2) \, \overline{\langle e_{i \lambda} | f \rangle} \langle e_{1 \lambda} | f \rangle\,,
\label{scadbro}
\end{equation}
\begin{equation}
\langle f | \ell \rangle_{\mathfrak{D}^{1/2}} :=
\langle f | \ell \rangle + \langle |\Hop|^{1/2} f \big| |\Hop|^{1/2} \ell \rangle
= (1 + |\mu_1|) \, \overline{\langle e_1 | f \rangle} \langle e_1 | \ell \rangle
+ \sum_{i=1}^2 \int_{0}^{+\infty} \hspace{-0.4cm} d \lambda \,
(1 + \lambda) \, \overline{\langle e_{i \lambda} | f \rangle} \langle e_{1 \lambda} | f \rangle\,.
\label{scad12bro}
\end{equation}
\textbf{Solution of the master equation.}
Let us consider Eq.~(\ref{masterbro}) with initial conditions~(\ref{cond}) i.e.,
$\ddot{\chi}(t) + \Hop \chi(t) = 0$, $\chi(0)=q$, $\dot{\chi}(0) = p$;
the unknown is a function $\Real \ni t \mapsto \chi(t) \in \mathfrak{D}$. We first
proceed formally, assuming that the initial data $q, p$ are in suitable spaces
to be specified later. Applying $\langle e_1 |~\rangle$ and
$\langle e_{i \lambda}|~\rangle$ to Eq.~(\ref{masterbro}) we obtain
$(d^2/ d t^2 + \mu_1) \langle e_1 | \chi(t) \rangle = 0$ and
$(d^2/ d t^2 + \lambda) \langle e_{i \lambda} | \chi(t) \rangle = 0$
for $i=1,2$ and all $\lambda > 0$. On account of the initial conditions
(\ref{cond}) (and recalling that $\mu_1 < 0$), these equations imply
\begin{equation}
\langle e_1 | \chi(t) \rangle = \langle e_1 | q \rangle \cosh( |\mu_1|^{1/2} t) +
\langle e_1 | p \rangle \frac{\sinh( |\mu_1|^{1/2} t)}{|\mu_1|^{1/2}} \,,
\quad \langle e_{i \lambda} | \chi(t) \rangle =
\langle e_{i \lambda} | q \rangle \cos( \lambda^{1/2} t) +
\langle e_{i \lambda} | p \rangle \frac{\sin( \lambda^{1/2} t)}{\lambda^{1/2}}\,,
\end{equation}
thus providing a formal justification for the expression~(\ref{solbro})
of $\chi(t)$. It can be checked a posteriori that, assuming
\begin{equation} q \in \mathfrak{D}\,, \quad p \in \mathfrak{D}^{1/2}\,,
\label{assuqp}
\end{equation}
all the previous manipulations make sense
and Eq.~(\ref{solbro}) describes the unique solution $\chi: \Real \ni t \mapsto \chi(t)$
of Eqs.~(\ref{masterbro},\ref{cond}) such that
\begin{equation}
\label{solspace}
\chi \in C^2(\Real, \Hilb) \cap C^1(\Real,
\mathfrak{D}^{1/2}) \cap C(\Real, \mathfrak{D})\,.
\end{equation}
As an example of the necessary tests, let us consider any $t \in \Real$ and show
that $\chi(t)$ defined by Eq.~(\ref{solbro}) is an element of $\mathfrak{D}$.
Due to the descriptions~(\ref{unit},\ref{unitinv}) for $\Hilb$ and
(\ref{reph}) for $\mathfrak{D}$, $\chi(t)$ in Eq.~(\ref{solbro}) is in fact in $\mathfrak{D}$ if
we are able to prove the following for $i=1,2$ (and for fixed $t$, as already indicated):
\\ \noindent
\vbox{
\begin{equation}
\lambda \mapsto \left[ \langle e_{i \lambda} | q \rangle \cos( \lambda^{1/2} t) +
\langle e_{i \lambda} | p \rangle \frac{\sin( \lambda^{1/2} t)}{\lambda^{1/2}} \right]
\in L^2((0,+\infty), d \lambda)\,,
\label{ver1}
\end{equation}
\begin{equation}
\lambda \mapsto \lambda \left[ \langle e_{i \lambda} | q \rangle \cos( \lambda^{1/2} t) +
\langle e_{i \lambda} | p \rangle \frac{\sin( \lambda^{1/2} t)}{\lambda^{1/2}} \right] =
\lambda \langle e_{i \lambda} | q \rangle \cos( \lambda^{1/2} t) +
\lambda^{1/2} \langle e_{i \lambda} | p \rangle \sin( \lambda^{1/2} t)
\in L^2((0,+\infty), d \lambda)\,.
\label{ver2}
\end{equation}
}
\noindent
Indeed, Eq.~(\ref{ver1}) follows noting that
\begin{equation}
\lambda \mapsto \cos( \lambda^{1/2} t)\,,~\lambda \mapsto \frac{\sin( \lambda^{1/2} t)}{\lambda^{1/2}}
\in L^\infty((0,+\infty), d \lambda)\,;\quad
\lambda \mapsto \langle e_{i \lambda} | q \rangle\,,~\lambda \mapsto \langle e_{i \lambda} | p \rangle
\in L^2((0,+\infty), d \lambda)
\label{a18}
\end{equation}
(the statements on $q, p$ in~(\ref{a18}) are correct, since Eq.~(\ref{assuqp}) obviously implies $q, p \in \Hilb$). Moreover, Eq.~(\ref{ver2}) follows noting that
\begin{equation}
\lambda \mapsto \cos( \lambda^{1/2} t)\,,~\lambda \mapsto \sin( \lambda^{1/2} t)
\in L^\infty((0,+\infty), d \lambda)\,;\quad
\lambda \mapsto \lambda \langle e_{i \lambda} | q \rangle\,,~\lambda \mapsto \lambda^{1/2}
\langle e_{i \lambda} | p \rangle
\in L^2((0,+\infty), d \lambda)
\label{a19} \end{equation}
(the statements on $q, p$ in~(\ref{a19}) are correct, due to the assumption~(\ref{assuqp}) that
$q \in \mathfrak{D}, p \in \mathfrak{D}^{1/2}$ and to the characterizations
(\ref{reph}) for $\mathfrak{D}$, and (\ref{reph12}) for $\mathfrak{D}^{1/2}$).

\section{On the master equation for the AdS wormhole}
\label{appeads}

\noindent
\textbf{Facts on the operator $\Hop$.}
Let us consider the Hilbert space of Eq.~(\ref{hilbu}), i.e.,
$\Hilb := L^2((-\pi,\pi), du)$, with its natural inner product $\langle~|~\rangle$;
this is the environment for
the selfajoint operator of Eq.~(\ref{hB}), i.e.,
$\Hop := -{d^2 \over du^2}+ \mathcal{ V} : \mathfrak{D} \subset \Hilb \to \Hilb$
with domain $\mathfrak{D} := \{ f \in \Hilb~|~f_{u u} \in \Hilb~, f(\pm \pi) = 0 \}$. Let us
recall that $\Hop$ has purely discrete spectrum with simple eigenvalues
$\mu_1 < 0 < \mu_2 < \mu_3 < \cdots$ (see Eq.~(\ref{signs})); due to Eq.~(\ref{Eq:boundsmun}), we have $\mu_n \sim {n^2 \over 4}$ for $n \mapsto + \infty$.

In the sequel we frequently make use of an orthonormal basis $(e_n)_{n=1,2,\ldots}$ of $\Hilb$ as in Eq.~(\ref{basisads}), obtained choosing for each $n$ a normalized eigenfunction
$e_n$ for the eigenvalue $\mu_n$. The fact that we have an orthonormal basis ensures that there is  a one-to-one linear map
\begin{equation}
\Hilb \to \mathfrak{l}^2\,, \quad f \mapsto \big(\langle e_n| f \rangle\big)_{n=1,2,\ldots}
\label{Eq:B1}
\end{equation}
where $\mathfrak{l}^2$ is the Hilbert space of complex sequences $(a_n)_{n=1,2\ldots}$
such that $\sum_{n=1}^{+\infty} |a_n|^2 < + \infty$, with its obvious inner product; moreover
$\langle f | \ell \rangle =
\sum_{n=1}^{+\infty} \overline{\langle e_n| f \rangle} \langle e_n| \ell \rangle$
for all $f, \ell \in \Hilb$, i.e., the map~(\ref{Eq:B1}) is unitary.

The fact that the orthonormal basis is formed by eigenfunctions of $\Hop$ ensures
the following representation for this operator and its domain
\begin{equation}
\mathfrak{D} = \{ f \in \Hilb~|~\big(\mu_n \langle e_{n}| f \rangle \big)_{n=1,2,\ldots}
\hspace{-0.2cm} \in \mathfrak{l}^2 \}\,.~
\mbox{For}~ f \in \mathfrak{D}:~\langle e_n | \Hop f \rangle = \mu_n
\langle e_{n}| f \rangle~(n=1,2,\ldots)\,,\mbox{i.e.}\,,~
\Hop f = \sum_{n=1}^{+\infty} \mu_n \langle e_n | f \rangle e_n\,.
\label{rephb}
\end{equation}
In the previous Appendix~\ref{appebro} we have already mentioned
the functional calculus for selfadjoint Hilbert space operators~\cite{Schmudgen-Book}; this allows to define an operator
$\mathscr{F}(\Hop) : \mathfrak{D}^{\mathscr{F}}
\subset \Hilb \to \Hilb$ for each function $\mathscr{F} :
\sigma(\Hop) \to \Complex$ where
$\sigma(\Hop) = \{ \mu_1, \mu_2,\ldots\}$ is the spectrum
of $\Hop$; the operator $\mathscr{F}(\Hop)$ is selfadjoint if
$\mathscr{F}$ is real valued. With the choice $\mathscr{F}(\gamma) := |\gamma|^{1/2}$
we obtain a selfajoint operator indicated with
\begin{equation}
| \Hop |^{1/2} : \mathfrak{D}^{1/2} \subset \Hilb \to \Hilb\,,
\end{equation}
which behaves as follows with respect to the previous orthonormal basis $(e_n)_{n=1,2,\ldots}$
of eigenfunctions of $\Hop$:
\begin{equation}
e_n \in \mathfrak{D}^{1/2}\,,\quad | \Hop |^{1/2} e_n = | \mu_n |^{1/2} e_n ~~ (n=1,2,\ldots)\,,
\end{equation}
\begin{equation}
\begin{gathered}
\mathfrak{D}^{1/2} = \{ f \in \Hilb~|~\big(|\mu_n|^{1/2} \langle e_{n}| f \rangle \big)_{n=1,2,\ldots}
\hspace{-0.2cm} \in \mathfrak{l}^2 \}\,. \\
\mbox{For}~ f \in \mathfrak{D}^{1/2}:~~\langle e_n | |\Hop|^{1/2} f \rangle = |\mu_n|^{1/2}
\langle e_{n}| f \rangle~(n=1,2,\ldots)\,,~~\mbox{i.e.}\,,~~
 | \Hop |^{1/2}  f = \sum_{n=1}^{+\infty} |\mu_n|^{1/2} \langle e_n | f \rangle e_n
\end{gathered}
\label{rephb12}
\end{equation}
(here and in the sequel, recall that $|\mu_n| = \mu_n$
for $n \geqslant 2$). Now, let us give Hilbert space structures to the domains $\mathfrak{D}$ and $\mathfrak{D}^{1/2}$;
these are provided by the (complete) inner products
$\langle~|~\rangle_{\mathfrak{D}} : \mathfrak{D} \times \mathfrak{D} \to \Complex$ and
$\langle~|~\rangle_{\mathfrak{D}^{1/2}} : \mathfrak{D}^{1/2} \times \mathfrak{D}^{1/2} \to \Complex$, where
 ({\footnote{One could as well consider the alternative inner products
$\langle~|~\rangle'_{\mathfrak{D}} : \mathfrak{D} \times \mathfrak{D} \to \Complex$ and
$\langle~|~\rangle'_{\mathfrak{D}^{1/2}} : \mathfrak{D}^{1/2} \times \mathfrak{D}^{1/2} \to \Complex$,
defined by setting
$$
\langle f | \ell \rangle'_{\mathfrak{D}} := \langle f | \ell \rangle
+ \langle \Hop f | \Hop \ell \rangle
= \sum_{n=1}^{+\infty} (1 + \mu^2_n) \, \overline{\langle e_n | f \rangle} \langle e_n | \ell \rangle\,,\qquad
\langle f | \ell \rangle'_{\mathfrak{D}^{1/2}} := \langle f | \ell \rangle
+ \langle |\Hop|^{1/2} f \big| |\Hop|^{1/2} \ell \rangle
= \sum_{n=1}^{+\infty} (1 + |\mu_n|) \, \overline{\langle e_n | f \rangle} \langle e_n | \ell \rangle\,;
$$
these have structures more similar to those of the inner products for the spaces $\mathfrak{D}$
and $\mathfrak{D}^{1/2}$ in the previous Appendix, see Eqs.~(\ref{scadbro},\ref{scad12bro}).
However, in the present situation $\langle~|~\rangle'_{\mathfrak{D}}$
and $\langle~|~\rangle'_{\mathfrak{D}^{1/2}}$ are equivalent, respectively,
to the inner products $\langle~|~\rangle_{\mathfrak{D}}$
and $\langle~|~\rangle_{\mathfrak{D}^{1/2}}$ of Eqs.~(\ref{scadads},\ref{scad12ads}).
}})
\begin{equation}
\langle f | \ell \rangle_{\mathfrak{D}} := \langle \Hop f | \Hop \ell \rangle
= \sum_{n=1}^{+\infty} \mu^2_n \, \overline{\langle e_n | f \rangle} \langle e_n | \ell \rangle\,,
\label{scadads}
\end{equation}
\begin{equation}
\langle f | \ell \rangle_{\mathfrak{D}^{1/2}} :=
\langle |\Hop|^{1/2} f \big| |\Hop|^{1/2} \ell \rangle
= \sum_{n=1}^{+\infty} |\mu_n| \, \overline{\langle e_n | f \rangle} \langle e_n | \ell \rangle\,.
\label{scad12ads}
\end{equation}
\textbf{Solution of the master equation.}
Let us consider the master equation as written in
Eq.~(\ref{masterads}) with the initial conditions given therein, i.e.,
$\ddot{\chi}(s) + \Hop \chi(s) = 0$, $\chi(0)=q$, $\dot{\chi}(0) = p$;
the unknown is a function $\Real \ni s \mapsto \chi(s) \in \mathfrak{D}$,
and the spaces containing the data $q, p$ are to be specified.
As in the previous Appendix we first
proceed formally. Applying $\langle e_n |~\rangle$ to the
differential equation in~(\ref{masterads}) we obtain
$(d^2/ d s^2 + \mu_n) \langle e_n | \chi(s) \rangle = 0$ for
$n=1,2,\ldots\,$; taking into account the initial conditions in
(\ref{masterads}) and the fact that $\mu_1 < 0 < \mu_2 < \mu_3 < \cdots$, we conclude that
\begin{equation}
\begin{gathered}
\langle e_1 | \chi(s) \rangle = \langle e_1 | q \rangle \cosh( |\mu_1|^{1/2} s) +
\langle e_1 | p \rangle \frac{\sinh( |\mu_1|^{1/2} s)}{|\mu_1|^{1/2}}\,, \\
\langle e_{n} | \chi(s) \rangle =
\langle e_{n } | q \rangle \cos( \mu^{1/2}_n s) +
\langle e_{n } | p \rangle \frac{\sin( \mu^{1/2}_n s)}{\mu^{1/2}_n} \quad (n=2,3,\ldots)\,,
\end{gathered}
\end{equation}
thus providing a formal justification for the expression~(\ref{solads})
of $\chi(s)$. It can be checked a posteriori that assuming
$q \in \mathfrak{D}$ and $p \in \mathfrak{D}^{1/2}$,
all the previous manipulations make sense
and Eq.~(\ref{solads}) describes the unique solution $\Real \ni s \mapsto \chi(s)$
of Eq.~(\ref{masterads}) in the space $C^2(\Real, \Hilb) \cap C^1(\Real,
\mathfrak{D}^{1/2}) \cap C(\Real, \mathfrak{D})$.
The verification of these statements relies on arguments similar to those
exemplified after Eqs.~(\ref{assuqp},\ref{solspace}) of the previous Appendix.

\bibliographystyle{unsrt}
\bibliography{refs_wormholes_rev}

\end{document}